\documentclass[sigconf]{acmart}

\usepackage{epstopdf}
\usepackage{algorithm}
\usepackage{multirow}
\usepackage{algorithmic}
\usepackage{color}
\usepackage[title]{appendix}
\usepackage{subfigure}
\usepackage{url}
\usepackage{xspace}
\usepackage{bbm}
\usepackage{fancyhdr}
\newcommand{\ie}{\emph{i.e.,}\xspace}
\newcommand{\eg}{\emph{e.g.,}\xspace}
\newcommand{\resp}{\emph{resp.,}\xspace}
\newcommand{\etal}{\emph{et al.}\xspace}
\newcommand{\eat}[1]{}
\def\EndOfProof{\nolinebreak\ \hfill\rule{1.5mm}{2.7mm}}

\makeatletter

\eat{

\newcommand{\SWITCH}[1]{\STATE \textbf{switch} #1\begin{ALC@g}}
\newcommand{\algorithmicendswitch}{\end{ALC@g}\textbf{end switch}}

\newcommand{\CASE}[2]{\STATE \textbf{case} #1\textbf{:}\begin{ALC@g}\STATE #2\end{ALC@g}}
\newcommand{\DEFAULT}{\textbf{default:}\begin{ALC@g}}
\newcommand{\ENDDEFAULT}{\end{ALC@g}}
}
\makeatother

\setcopyright{none}

\begin{document}
\fancyhead{}
\title{Towards Enhancing Database Education: Natural Language Generation Meets Query Execution Plans}
\subtitle{[Technical Report]}
\author{Weiguo Wang $^{\ddag,\S}$ \hspace{5ex} Sourav S Bhowmick$^\ddag$ \hspace{5ex} Hui Li$^\S$ 
\hspace{5ex} Shafiq R Joty$^\ddag$ \\ \hspace{5ex} Siyuan Liu $^{\ddag}$ \hspace{5ex} Peng Chen$^\S$ }
\affiliation{
  $^\ddag$School of Computer Science and Engineering, Nanyang Technological University, Singapore\\
  $^\S$School of Cyber Engineering, Xidian University, China
  }
\email{assourav|srjoty|sliu019@ntu.edu.sg, hli@xidian.edu.cn, wgwang|pchen97@stu.xidian.edu.cn}
\eat{\renewcommand{\shortauthors}{Z. Ye et al.}}

\begin{abstract}
The database systems course is offered as part of an undergraduate computer science degree program in many major universities. 
A key learning goal of learners taking such a course is to understand how \textsc{sql} queries are processed in a \textsc{rdbms} in \emph{practice}.  Since a \textit{query execution plan} (\textsc{qep}) describes the execution steps of a query, learners can acquire the understanding by perusing the \textsc{qep}s generated by a \textsc{rdbms}. Unfortunately, in practice, it is often daunting for a learner to comprehend these  \textsc{qep}s containing vendor-specific implementation details, hindering her learning process. In this paper, we present a novel, end-to-end, \textit{generic} system called \textsc{lantern} that generates a natural language description of a \textsc{qep} to facilitate understanding of the query execution steps. It takes as input an \textsc{sql} query and its \textsc{qep}, and generates a  natural language description of the execution strategy deployed by the underlying \textsc{rdbms}. Specifically, it deploys a \textit{declarative framework} called \textsc{pool} that enables \textit{subject matter experts} to efficiently create and maintain natural language descriptions of physical operators used in \textsc{qep}s. A \textit{rule-based} framework called \textsc{rule-lantern} is proposed that exploits \textsc{pool} to generate natural language descriptions of \textsc{qep}s.
Despite the high accuracy of \textsc{rule-lantern}, our engagement with learners reveal that, consistent with existing psychology theories, perusing such rule-based descriptions lead to \textit{boredom} due to repetitive statements across different \textsc{qep}s.  To address this issue, we present a novel \textit{deep learning-based} language generation framework called \textsc{neural}-\textsc{lantern} that infuses language variability in the generated description by exploiting a set of \textit{paraphrasing tools} and \textit{word embedding}.  Our experimental study with real learners shows the effectiveness of \textsc{lantern} in facilitating comprehension of \textsc{qep}s.
\end{abstract}

%
%



\maketitle

\vspace{0ex}
\section{Introduction}\label{sec:intro}
There is continuous demand for database-literate professionals in today's market due to the widespread usage of relational database management system (\textsc{rdbms}) in the commercial world. Such commercial demand has played a pivotal role in the offering of database systems course as part of an undergraduate computer science (\textsc{cs}) degree program in major universities around the world. Furthermore, not all working adults dealing with \textsc{rdbms} have taken an undergraduate database course. Hence, they often need to undergo on-the-job training or attend relevant courses in higher institutes of learning to acquire database literacy. Indeed, while formal education for young learners at universities has been the focus of educational provisions in the industrial age, the digital age is now seeing an increased experimentation of  ``lifelong learning''~\cite{unesco} with provisions such as  work-study programmes for early career and mid-career individuals, and digital learning initiatives.

A key learning goal of learners taking a database course is to understand how \textsc{sql} queries are processed in a \textsc{rdbms} in practice.  A relational query engine produces a \textit{query execution plan} (\textsc{qep}), which represents an execution strategy of an \textsc{sql} query.  Hence, such understanding can be gained by learners by perusing the \textsc{qep}s of queries. Major database textbooks introduce the \emph{general} (\ie  not tied to any specific \textsc{rdbms} software) theories and principles behind the generation of \textsc{qep}s using natural language-based narratives and visual examples. This allows a learner to gain a general  understanding of query execution strategies of \textsc{sql} queries. 

\eat{Hence, a learner typically studies the \textsc{qep} of an \textsc{sql} query to gain understanding of its execution.}

{Most database courses complement text book-learning  with hands-on interaction with an off-the-shelf commercial \textsc{rdbms} (\eg PostgreSQL) to infuse knowledge about database techniques used in \textit{practice}. A learner will typically implement a database application, pose queries over it, and peruse the associated \textsc{qep}s to comprehend how they are processed by a commercial-grade query engine. Most commercial \textsc{rdbms} expose the \textsc{qep} of an \textsc{sql} query using \textit{visual} or \textit{textual} (\eg unstructured text, \textsc{json}, \textsc{xml}) format.  Unfortunately, comprehending these textual formats to understand query execution strategies of \textsc{sql} queries in practice is daunting for learners. In contrast to natural language-based narrations in database textbooks, they are not user-friendly and assume deep knowledge of vendor-specific implementation details.  On the other hand, the visual format is relatively more user-friendly but hides important details. Consequently, it is challenging for learners to  understand query execution strategies in a specific \textsc{rdbms} from these \textsc{qep} formats. We advocate that in order to promote palatable learning experiences for diverse individuals in full recognition of the complexity of \textsc{qep}s in practice, user-friendly tools are paramount.

\eat{\begin{figure}[t]
\centering
\includegraphics[width=0.8\linewidth,height=4.5cm]{sql.eps}
\vspace{-2ex}\caption{Query 4 in \textsc{tpc-h} benchmark dataset.}
\label{fig:query}
\vspace{-2ex}\end{figure}}

\eat{
\begin{figure}[t]
\centering
\includegraphics[width=\linewidth]{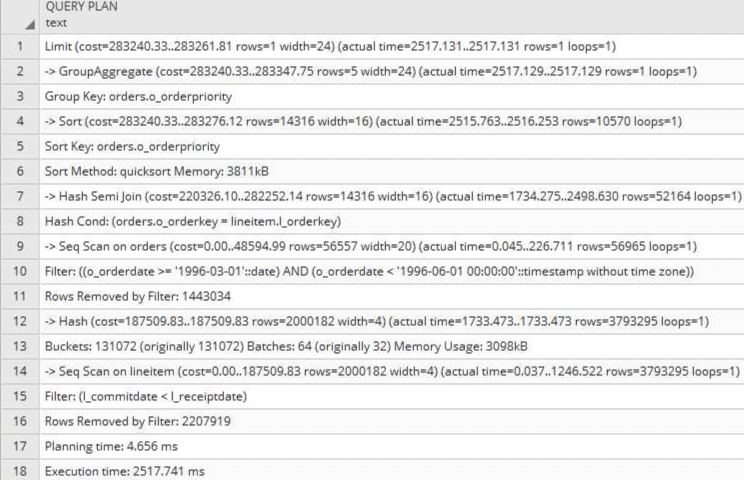}
\vspace{-3ex}\caption{A \textsc{qep} in PostgreSQL.}
\label{fig:plan1}
\vspace{-2ex}\end{figure}

\begin{figure}[t]
\centering
\includegraphics[width=\linewidth]{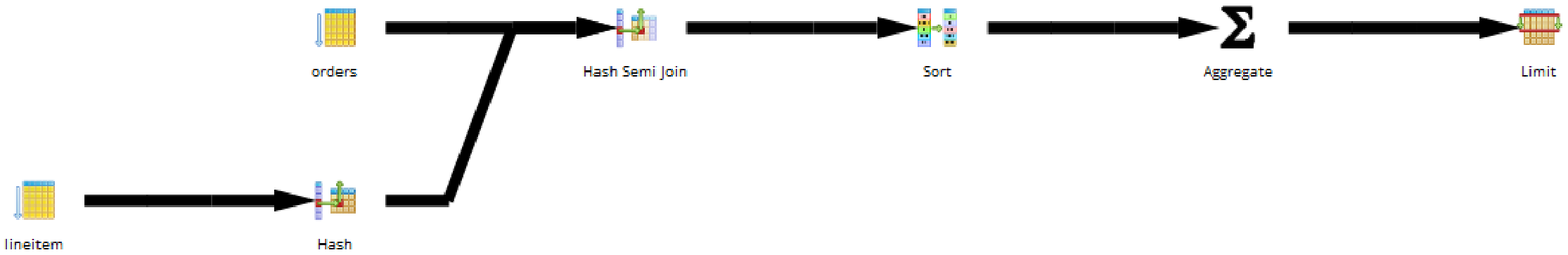}
\vspace{-3ex}\caption{Visual tree representation of the \textsc{qep}.}
\label{fig:plan2}
\vspace{-3ex}\end{figure}}

\begin{figure}[t]
\centering
\includegraphics[width=\linewidth]{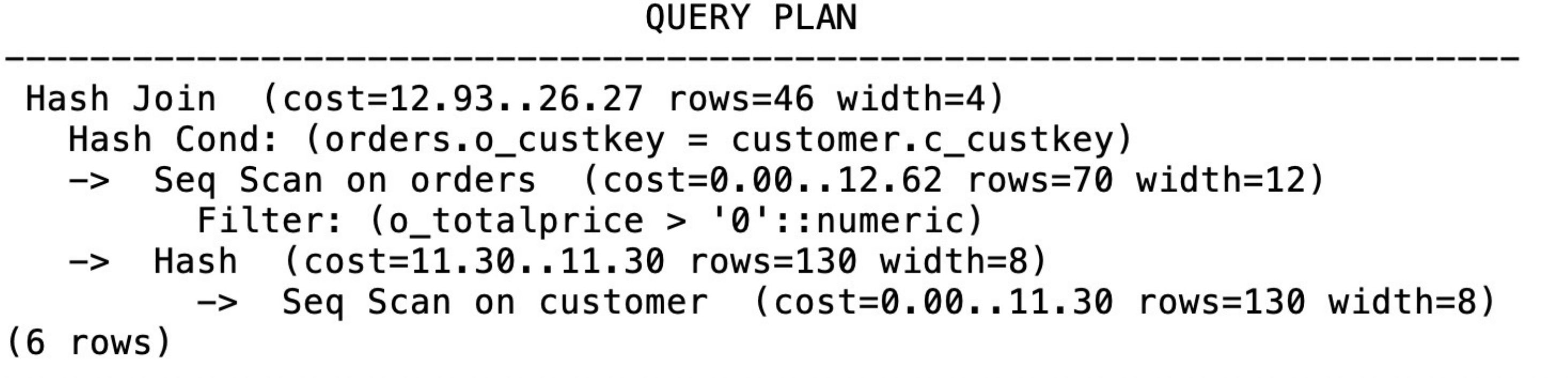}
\vspace{-4ex}\caption{A \textsc{qep} in PostgreSQL.}
\label{fig:plan1}
\vspace{-3ex}\end{figure}

\begin{figure}[t]
\centering
\includegraphics[width=\linewidth]{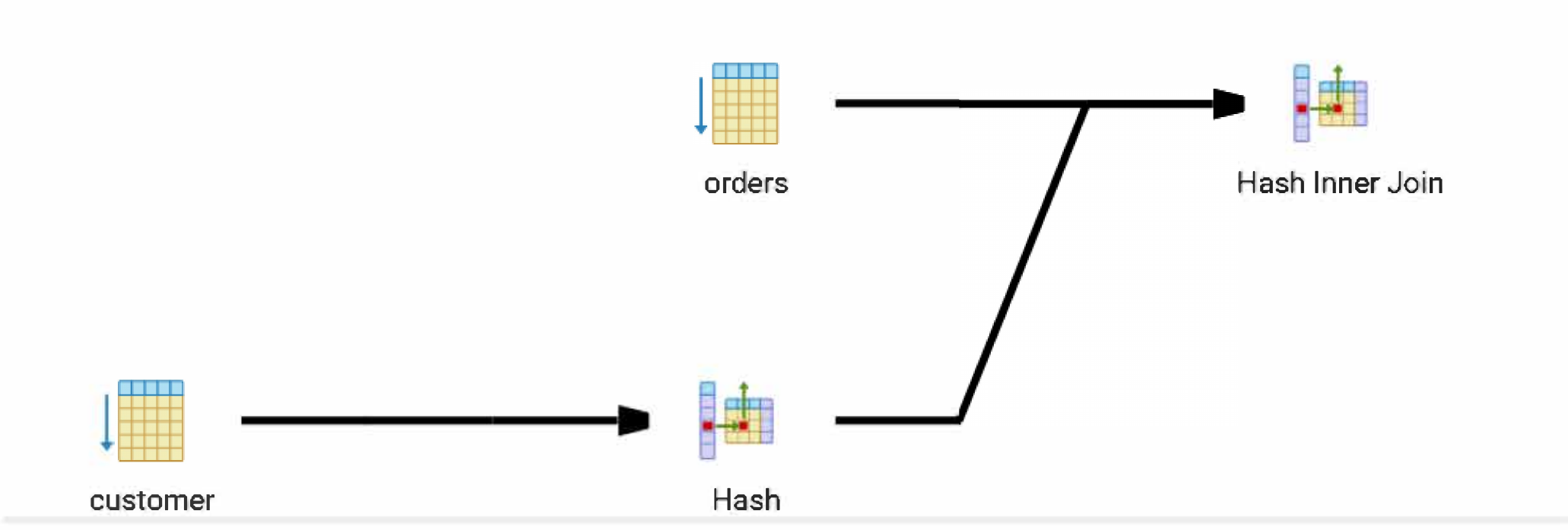}
\vspace{-3ex}\caption{Visual tree representation of the \textsc{qep}.}
\label{fig:plan2}
\vspace{-3ex}\end{figure}

\begin{example}
Alice is an undergraduate \textsc{cs} student who is currently enrolled in a database course. She wishes to understand the execution steps of an \textsc{sql} query in PostgreSQL on a \textsc{tpc-h} benchmark dataset~\cite{tpch} by perusing the corresponding \textsc{qep} in Figure~\ref{fig:plan1} (partial view). Unfortunately, Alice finds it difficult  to mentally construct a narrative of the overall execution steps by simply perusing it. This problem is further aggravated in more complex \textsc{sql} queries. Hence, she switches to the visual tree representation of the \textsc{qep} as shown in Figure~\ref{fig:plan2}. Although relatively succinct, it simply depicts the sequence of operators used for processing the query, hiding additional details about the query execution (\eg sequential scan, join conditions).  In fact, Alice needs to manually delve into details associated with each node in the tree for further information.
\EndOfProof\end{example}

We advocate that an  intuitive natural language-based description of a \textsc{qep} can greatly facilitate  learners to comprehend how an \textsc{sql} query is executed by a \textsc{rdbms}. To support this hypothesis, we surveyed 62 unpaid volunteers taking the database course in an undergraduate \textsc{cs} degree program.  We use the \textsc{tpc-h} v2.17.3 benchmark  and a  rule-based natural language generation tool for \textsc{qep}s ~\cite{neuron} to generate natural language  (\textsc{nl}) descriptions of \textsc{qep}s for \textsc{sql} queries formulated by the volunteers (both ad hoc and benchmark queries).  The volunteers were asked to select their most preferred \textsc{qep} format (\ie \textsc{json} text, visual tree, and \textsc{nl} description)  that aide in understanding the execution steps of these queries. Figure~\ref{fig:survey} depicts the results. Observe that \textsc{nl} description is the most preferred format. On the other hand, very few voted for the \textsc{json} format supported by  PostgreSQL. Also, the visual tree representation of a \textsc{qep} has healthy support. Hence, we believe that an \textsc{nl}-based interface can effectively \textit{complement} visual \textsc{qep}s to augment the learning experience of learners. Specifically, a learner may use the visual \textsc{qep} to get a quick overview of an execution plan and then peruse the \textsc{nl} description to acquire detailed understanding.

The majority of natural-language interfaces for \textsc{rdbms}~\cite{LJ14a,LJ14b,SF+16,KV+12}, however, have focused either on translating natural language sentences to \textsc{sql} queries or narrating \textsc{sql} queries in a natural language. Scant attention has been paid for generating natural language descriptions of \textsc{qep}s. Natural language generation for \textsc{qep}s is challenging from several fronts. First, although deep learning techniques, which can learn task-specific representation of input data, are particularly effective for natural language processing, it has a major upfront cost. These techniques need massive training sets  of labeled examples to learn from. Such training sets in our context are prohibitively expensive to create as they demand database experts to translate thousands of \textsc{qep}s of a wide variety of \textsc{sql} queries.  Even  labeling using crowdsourcing is challenging as accurate natural language descriptions demand experts who understand \textsc{qep}s. Note that accuracy is critical here as low quality translation may adversely impact individuals' learning.\eat{ Furthermore, rules and heuristics for labeling a \textsc{qep} impact the accuracy and coverage due to wide variety of \textsc{qep}s a query optimizer generates.} Second, ideally we would like to generate natural language descriptions of \textsc{qep}s using one application-specific dataset (\eg movies) and then use it for other applications (\eg hospital). That is, the natural language generation framework should be \textit{generalizable}. This will significantly reduce the cost of its deployment in different learning institutes and environments where different application-specific examples may be used to teach database systems.

In this paper, we present a novel end-to-end system called \textsc{lantern} (natura\underline{L} l\underline{AN}guage descrip\underline{T}ion of qu\underline{ER}y pla\underline{N}s) to generate natural-language descriptions of \textsc{qep}s. Given an \textsc{sql} query and its \textsc{qep}, it automatically generates a natural language description of the key steps undertaken by the underlying \textsc{rdbms} to execute the query.  To this end, instead of  mapping an \textit{entire} \textsc{qep} to its natural language description, we focus on mapping the set of physical operators in a \textsc{rdbms} to corresponding natural language descriptions and then \textit{stitch} them together to generate the description of a specific \textsc{qep}.   The rationale behind this strategy is as follows. Any \textsc{rdbms} implements a small number of physical operators to execute any \textsc{sql} query. Hence, although there can be numerous \textsc{qep}s, they are all built from a small set of physical operators. Consequently, it is more manageable to label these operators and generate the natural language description of any \textsc{qep} from them. This also allows us to generalize \textsc{lantern} to handle any application-specific database as the relations, attributes, and predicates can simply be used as placeholders in describing a physical operator. Lastly, it makes our framework \textit{orthogonal} to the complexities of \textsc{sql} queries as they are all executed by a small set of physical operators.

\begin{figure}[t]
\centering
\includegraphics[width=0.9\linewidth]{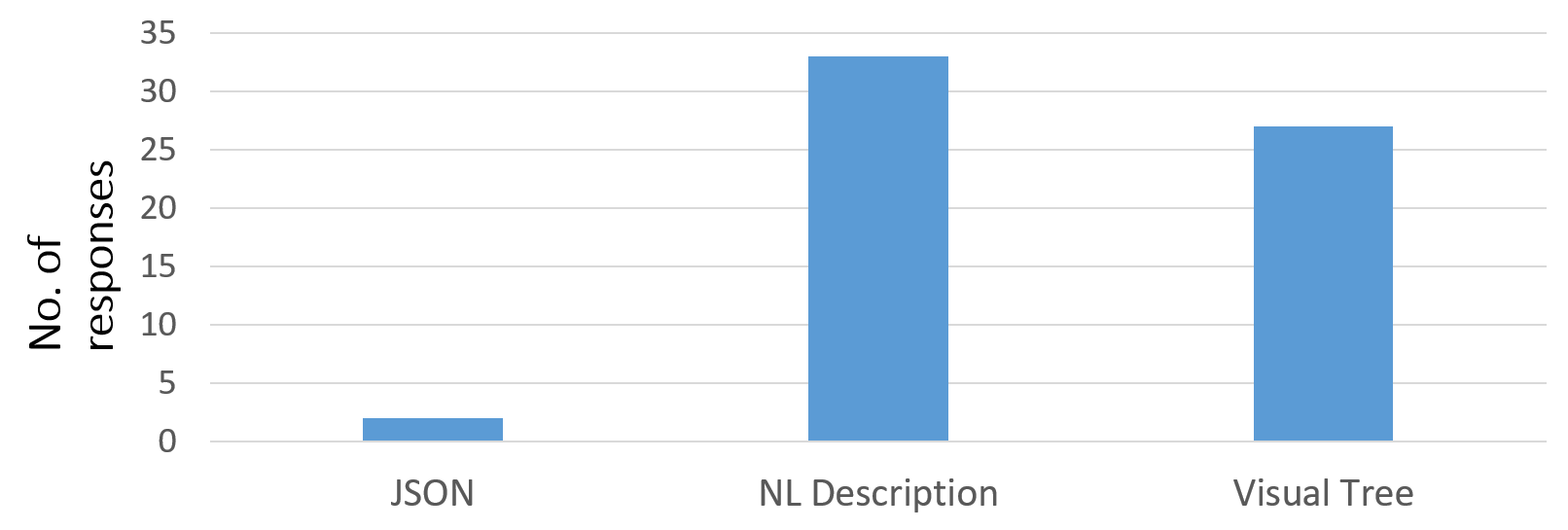}
\vspace{-3ex}\caption{Survey of \textsc{qep} formats.}
\label{fig:survey}
\vspace{0ex}\end{figure}

We present a  flexible \textit{declarative} framework called \textsc{pool} for succinctly specifying natural language descriptions of physical operators in an \textsc{rdbms}. We then develop a \textit{rule-based} framework called \textit{\textsc{rule}-\textsc{lantern}} to generate a natural language description  of a \textsc{qep} by leveraging the specified descriptions of physical operators. We observe from our engagements with learners that although  rule-based approach have high accuracy, it makes the descriptions of \textsc{qep}s monotonous leading to boredom. In fact, this is consistent with psychology theories that repetition of messages can lead to annoyance and boredom~\cite{CP79} (detailed in Section~\ref{sec:psycho}). To address this issue, we develop a novel \textit{deep learning-based} language generation framework called \textit{\textsc{neural}-\textsc{lantern}} that infuses language variability in the generated description by exploiting a group of \textit{paraphrasing tools}~\cite{synonymous1,synonymous2,synonymous3} and pretrained word embeddings~\cite{devlin2018bert,mikolov2013efficient,pennington2014glove,ELMO}. Importantly, it addresses the challenge of training data generation  by first generating a  large number of random queries based on schema information and actual values in the database and then utilize \textsc{rule}-\textsc{lantern} and the paraphrasing tools to generate a large number of natural language descriptions of the  physical operators. We built \textsc{lantern} on top of PostgreSQL and SQL Server. Our exhaustive experimental study with real learners demonstrates the superiority of \textsc{lantern} to existing \textsc{qep} formats of commercial \textsc{rdbms}.

In summary, this paper makes the following contributions: 

\begin{itemize}
\item We present a novel end-to-end system called \textsc{lantern} for generating natural language descriptions of \textsc{qep}s. It  takes a concrete step towards the vision of natural language interaction with the relational query optimizer. 

\item We present a \textit{declarative} framework called \textsc{pool} to enable \textit{subject matter experts}  (\textsc{sme}s) to label physical operators in an intuitive way (Section~\ref{sec:declare}). 

\item Based on the specifications using \textsc{pool}, in Section~\ref{sec:rule} we present a rule-based approach called \textsc{rule}-\textsc{lantern} to generate a natural language description of a \textsc{qep}.

\item  We present a novel psychology-inspired neural framework  for natural language generation called \textsc{neural}-\textsc{lantern} in Section~\ref{sec:neural} that addresses  limitations of \textsc{rule}-\textsc{lantern}. (e) In Section~\ref{sec:perf}, we undertake an exhaustive performance  study using synthetic and real-world datasets to demonstrate the effectiveness of \textsc{lantern}.
 
 \end{itemize}

\vspace{0ex}
\section{Related Work}
\eat{Query optimizers have been extensively studied since the inception of relational databases.  However, to the best of our knowledge, there has been scant research on natural language interaction with relational query optimizers.}

Natural language interfaces to databases have been studied for several decades. Such interfaces enable users easy access to data, without the need to learn a complex query languages, such as \textsc{sql}. Specifically, there have been natural language interfaces for relational databases~\cite{ASB19},\eat{ video databases~\cite{ECC08},} \textsc{xml}~\cite{LC+07}, and graph-structured data~\cite{ZC+17}. Given a logically complex English language sentence as query input, the goal of majority of these work is to translate it to the underlying query language such as \textsc{sql}~\cite{BJL19,BH+18,SF+16,Yl+16,YZ+18,YZ+19,BJ+20,WU19,ZXS,XC+19,KS+20}. Recently, deep learning techniques have been utilized to translate natural language queries to \textsc{sql}~\cite{BH+18,UW+18,XC+19,Yl+16,YZ+18,YZ+19,ZXS}. On the other hand, frameworks such as Logos~\cite{KV+12} explain \textsc{sql} queries to users using natural language. \textsc{lantern} compliments these efforts by providing a natural language description of a \textsc{qep}.

Most germane to this work is the demonstration of a system called \textsc{neuron} in~\cite{neuron}, which generates natural language descriptions of \textsc{qep}s using a rule-based technique. It also supports a \textit{natural language question answering}  system that allows a user to seek answers to a variety of concepts and features associated with a \textsc{qep}. In contrast, we focus on generating a natural language description of a \textsc{qep} and give detailed methodology to address this problem. We also introduce a \textit{declarative framework} for label specification and a deep learning-based solution that are omitted in~\cite{neuron}. Finally, user studies and experiments demonstrating the effectiveness of the proposed frameworks are presented in this work.

\begin{figure}[t]
\centering
\includegraphics[width=0.7\linewidth]{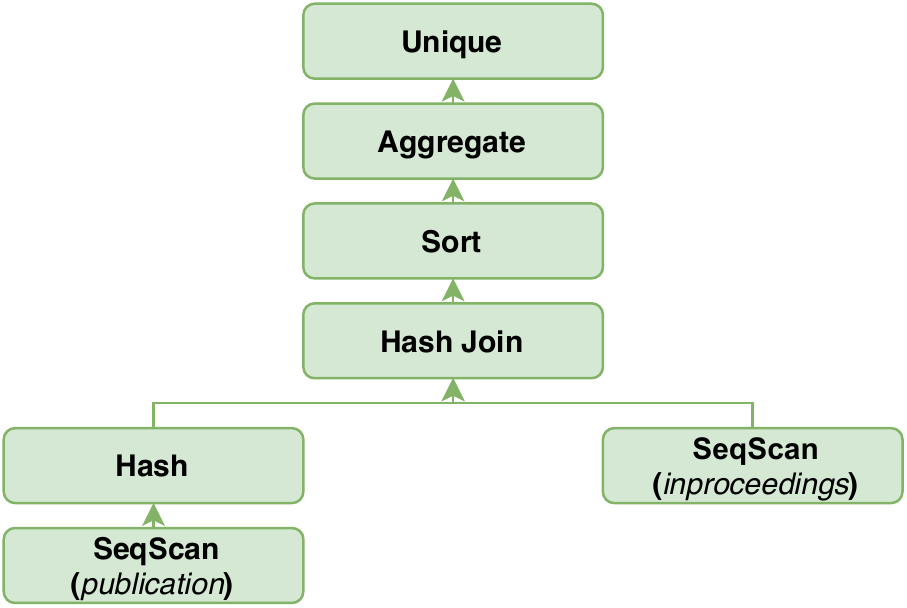}
\vspace{-2ex}\caption{A physical operator tree (\ie \textsc{qep}).}
\label{fig:exp-qep}
\vspace{0ex}\end{figure}

\vspace{0ex}
\section{Preliminaries} \label{sec:prelim}
In this section, we present basic concepts that are necessary to comprehend this paper. \eat{We begin with the concept of physical operator trees  in a relational query optimizer. Next, we briefly introduce the \textit{Seq2Seq} model used by \textsc{neural-lantern}.}

\vspace{1ex}
\noindent\textbf{ Physical Operator Tree.}
The relational query execution engine implements a set of physical operators~\cite{SC98}. An operator takes as input one or more data streams and produces an output data stream. Some examples of physical operators are sequential scan, index scan, and hash join. Note that in database literature, we typically refer to such operators as physical operators since there is not necessarily one-to-one mapping with relational operators. These  physical operators are the building blocks for the execution of \textsc{sql} queries. An abstract representation of such an execution is a\textit{ physical operator tree} (operator tree for brevity), denoted as $T=(N, E)$,  where $N$ is a set of nodes representing the operators and $E$ is a set of edges representing  data flow among the physical operators. The physical operator tree is the abstract representation of a query execution plan (\textsc{qep}) and we use these terms interchangeably. The query execution engine is responsible for the execution of the \textsc{qep} to generate results of a \textsc{sql} query.

\begin{example} \label{eg:1}
Consider the following \textsc{sql} query on \textsc{dblp} dataset.
\begin{quote}
\begin{verbatim}
SELECT DISTINCT(I.proceeding_key)
FROM inproceedings I, publication P
WHERE (I.proceeding_key = P.pub_key AND
       P.title like '%July%')
GROUP BY I.proceeding_key
HAVING COUNT (*) > 200;
\end{verbatim}
\end{quote}

The \textsc{qep} (\ie physical operator tree) of the query  generated by PostgreSQL is depicted in Figure~\ref{fig:exp-qep}.
\EndOfProof\end{example}

\eat{\begin{figure}[t]
	\centerline{\includegraphics[width=0.9\linewidth]{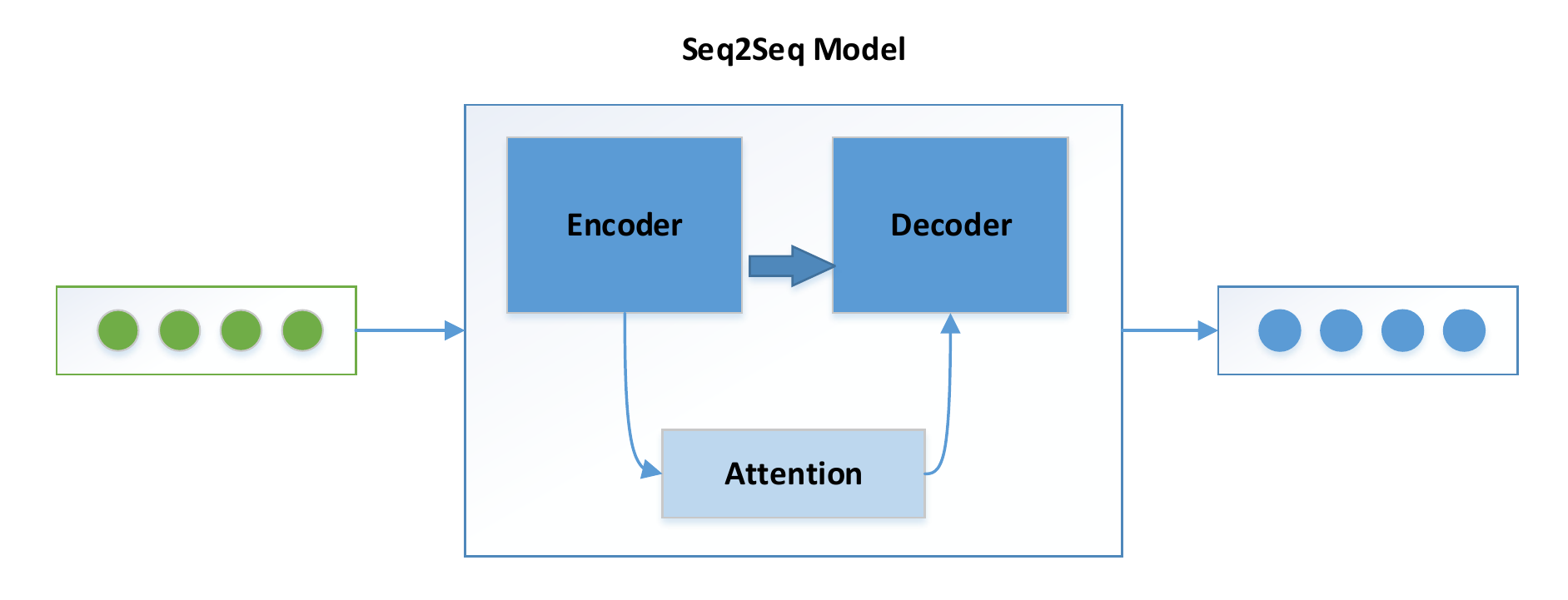}}
	\vspace{-3ex}\caption{Seq2Seq model.}
	\label{seq2seq}
\vspace{-3ex}\end{figure}}

Intuitively, we classify the nodes in an operator tree  of a \textsc{qep} into two categories, namely \textit{critical} and \textit{auxiliary} operators (nodes). The former type of nodes corresponds to important operations (\eg \textsf{HASH JOIN}, \textsf{SEQSCAN}) in a \textsc{qep}. On the other hand, the latter type is located near a critical node (\eg parent, child) and supports the execution of the operator represented by a critical node.\eat{ In fact, they are only interpreted in the context of critical nodes.} For instance, \textsf{HASH JOIN}  and \textsf{HASH} in Figure~\ref{fig:exp-qep} are examples of critical and auxiliary nodes/operators in PostgreSQL, respectively .\eat{ Similarly, in DB2, \textsf{TBSCAN} node and its child \textsf{FILTER} represent these two categories, respectively.}

\vspace{1ex}
\noindent\textbf{Seq2Seq Model.}
The\textit{ Seq2Seq} model has revolutionized the process of machine translation using the deep learning framework~\cite{sutskever2014sequence}. It has also been a standard method for other text generation applications such as image captioning \cite{Vinyals-caption,Karpathy-image-cap}, conversational models \cite{Vinyals-dialog}, text summarization \cite{see-summ}. The Seq2Seq model puts two neural networks together, one as an encoder and the other as a decoder. It takes as input a sequence of words and generates an output sequence of words. Given a source (input) sequence of words, the encoder-decoder framework works as follows. The \textit{encoder} deep neural network converts the input words to its corresponding hidden vectors, where each vector gives a contextual representation of the corresponding word. The \textit{decoder} deep neural network is a language model that takes as input the hidden vector generated by the encoder, its own hidden (previous) states and the current word to produce the next hidden vector and to finally predict the next word.

The encoder and the decoder can use a recurrent neural network (\textsc{rnn}) architecture \cite{sutskever2014sequence} with carefully designed cells like \textsc{lstm} or \textsc{gru}, convolutional \cite{convseq2seq,Wu-dynconv} or recently proposed transformer architecture \cite{Vaswani-transformer}. It is also common to employ an attention mechanism \cite{Bahdanau-attn,luong2015effective} so that the decoder can selectively focus on relevant encoder states while generating a token.

\vspace{0ex}
\section{A Declarative Framework} \label{sec:declare}
Ideally, we would like to have access to large volumes of labels that associate \textsc{qep}s to their corresponding natural language descriptions. Then, in principle, we can use such labels as training data to build a deep learning-based model to generate a natural language description of any \textsc{qep}. Unfortunately, although there is increasing availability of \textsc{sql}-to-natural language training datasets~\cite{YZ+18}, to the best of our knowledge, no publicly-available data source exists for \textsc{qep}s.  Note that natural language translation techniques for \textsc{sql} queries cannot be adopted here as \textsc{sql} queries are declarative and specified using logical operators.   To further aggravate this issue, the natural language labeling of \textsc{qep}s needs to be performed by trained \textit{subject matter experts} (\textsc{sme}) in order to ensure accuracy of the annotations.  Given that there can be numerous \textsc{qep}s in practice, it is prohibitively expensive to deploy such experts  for labeling \textsc{qep}s.  On the other hand, crowdsourcing for cheaper sources of labeling is not a viable option as non-\textsc{sme}s may not have sufficient background on query processing to annotate \textsc{qep}s with high degree of accuracy.

To address these challenges, we deploy \textsc{sme}s to provide natural language descriptions for physical operators in a commercial \textsc{rdbms} in lieu of \textsc{qep}s. All \textsc{qep}s are essentially constructed from this set of operators, which is orders of magnitude smaller than a training set containing \textsc{qep}s, making natural language descriptions (\ie labels) economical to obtain from \textsc{sme}s. In order to expedite the labeling process, we propose a \textit{declarative framework}  where a \textsc{sme} can create and manipulate the labels using a declarative language called \textsc{pool} (\underline{P}hysical \underline{O}perator \underline{O}bject \underline{L}anguage). In this section, we elaborate on this framework. Note that we focus on features that are necessary to understand our \textsc{rule-lantern} framework.\eat{ Exhaustive discussion of \textsc{pool} is beyond the scope of this work.}

\vspace{0ex}
\subsection{Requirements}
Investigation of physical operators in commercial \textsc{rdbms} as well as our engagement with learners identified several crucial requirements for \textsc{pool}. First, an abstract data type for physical operators is necessary so that \textsc{sme}s may treat such data at a level independent from a specific \textsc{rdbms}.

Second,  \textsc{sme}s must be able to select objects to be labeled  and specify corresponding natural language labels. In particular, they must be able to label physical operators in three dimensions, namely, create meaningful \textit{alias} of a physical operator, natural language \textit{definition} of an operator, and natural language \textit{descriptions} of the operation performed by an operator. In particular, aliases are important as names of certain operators can be ambiguous to a learner. For instance, DB2 uses \textsf{HSJOIN} as the name of hash join operator. Hence, an alias of  \textsf{HASH JOIN} is more intuitive to a learner. Similarly, a learner may encounter unfamiliar operators (\eg zigzag join (\textsf{ZZJOIN} in DB2)) in her course. Hence, natural language definitions of such operators will be useful to her while perusing \textsc{qep}s.

Third, it may be necessary to declaratively \textit{combine} labels of a pair of operators in order to generate a succinct natural language description of a \textsc{qep}. For instance, in PostgreSQL labels of \textsf{HASH JOIN}  and \textsf{HASH} operators need to be combined to generate a natural language description of the former operator. An example of such description can be  \textit{``hash \$$R_1$\$ and perform hash join on \$$R_2$\$ and \$$R_1$\$ on condition \$$cond$\$''} where  \textit{``hash \$$R_1$\$''} is the label associated with the \textsf{HASH} operator.

Fourth, \textsc{sme}s should be able to \textit{transfer} the description of one operator to another to make specification of natural language descriptions of operators more efficient. For instance, hash join and nested-loop join are both join operators. Consequently, their descriptions may be very similar, \ie the description of nested-loop join operator does not have  the  \textit{``hash \$$R_1$\$''} segment. Hence, \textsc{pool} should be able to reuse and modify the existing description of hash join operator when specifying the description of nested-loop join operator. Similarly, one should be able to transfer the description or definition of hash join across \textit{different} \textsc{rdbms} (\eg PostgreSQL to DB2) without specifying it from scratch.

\vspace{0ex}
\subsection{Features of POOL}

\textbf{Data Model.}
The data model underlying \textsc{pool} is called  \textsc{poem} (\underline{P}hysical \underline{O}perator Obj\underline{E}ct \underline{M}odel).  \textsc{poem} is a simple and flexible graph model where  all entities are objects. Each object represents a physical operator of a relational query engine. Each object has a unique \textit{object identifier} ($oid$) from the type \textit{oid}.  Objects are either atomic or complex.  Atomic objects do not have any outgoing edges. All objects have a set of attribute-value pairs. Specifically, each object is associated with the following attributes: \textit{source}, \textit{name}, \textit{alias}, \textit{defn}, \textit{desc}, \textit{type}, \textit{cond}, and \textit{target}. The \textit{source} refers to the specific relational engine that an operator belongs to. The \textit{name} refers to the name of a physical operator in the source and the \textit{type} captures whether it is an unary or binary operator. \textit{Alias} is an optional alternative name for an operator. The \textit{defn} attribute stores the definition of an operator. The \textit{desc} attribute stores a natural language description of the operation performed by an operator.  The \textit{cond} attribute takes a Boolean value to indicate whether a specified condition (\eg join condition) should be appended to the natural language description of an operator. Values of all attributes are taken from the atomic type \textsf{string} (possibly empty).  As an example, consider the \textsf{HASHJOIN} operator in PostgreSQL. In \textsc{poem}, it is an object with the following attributes: \textsf{source = `postgresql'}, \textsf{name = `hashjoin'}, \textsf{alias = ` '},  \textsf{type = `binary'}, \textsf{defn = `a type of join algorithm that uses hashing to create subsets of tuples'}, \textsf{desc = `perform hash join'}, and \textsf{cond = `true'}.  Note that the \textit{source} serves as an entry point to the database.  The set of objects is referred to as \textit{\textsc{poem} store}.

There is a directed edge between an object pair ($p_a$, $p_c$) iff $p_a$ is an auxiliary operator and $p_c$ is a critical operator (recall from Section~\ref{sec:prelim}) in a \textsc{qep} in \textit{source}. It is captured by the \textit{target} attribute of $p_a$. For example, ($p_\textsf{hash}$, $p_\textsf{hashjoin}$)  of PostgreSQL has a directed link. Hence, the \textit{target} attribute value of $p_\textsf{hash}$ is \textsf{`hashjoin}'.

\vspace{1ex}\noindent\textbf{Data Definition.}
The data definition in \textsc{pool} allows one to declaratively create physical operator objects associated with a specific \textsc{rdbms}. The general format of the statement is as follows: \textsf{CREATE POPERATOR} \textit{<name>} \textsf{FOR} \textit{<source>} (\textit{<attribute-value pairs>}). An example is as follows.

\begin{quote}
\vspace{-1ex}\begin{verbatim}
CREATE POPERATOR hashjoin FOR pg
(ALIAS = null,
TYPE = 'binary',
DEFN = null,
DESC = 'perform hash join',
COND = 'true',
TARGET = null)
\end{verbatim}
\end{quote}

Note that \textit{name} must exists in the set of physical operators supported by the specified \textsc{rdbms} engine (\ie source). For instance, \textsf{hashjoin} is a physical operator in PostgreSQL (\ie \textsf{pg}). The optional \textsf{ALIAS} attribute specifies an alternative name of the operator. For example, the operator \textsf{ZZJOIN} in DB2 can be given an alias \textsf{`zigzag join'} .  In the case an alias is unspecified, it can be either set to null or simply omitted from the definition.  The \textsf{TYPE} is a mandatory attribute that can take either \textsf{`unary'} or \textsf{`binary'} value. The \textsf{DESC} attribute is mandatory, which allows one to specify a natural language description of the operation performed by the operator.  Note that \textsc{pool} does not prevent one from describing several descriptions for a single operator. For instance, \textsf{DESC = `execute hash join'} can be added to the above definition. Observe that no relation or condition is specified in \textsf{DESC}. This is because these are added automatically to \textsf{DESC} by exploiting \textsf{TYPE} and \textsf{COND} attributes  of an operator. For instance, since \textsf{TYPE = `binary'} in the above definition, two variables representing join relations will be added automatically to the description of \textsf{hashjoin}.  Lastly, the \textsf{TARGET} attribute allows one to specify auxiliary-critical operator pair. If its value is non-empty, it must be an existing name in the source.

\vspace{1ex}\noindent\textbf{Data Manipulation.}
The key goals of the data manipulation component of \textsc{pool} are to provide syntactical means to support (a) retrieval of specific properties (\ie attributes) of physical operators, (b) generation of the \textit{template} for natural language description of  an operator that can be subsequently used in a \textsc{qep}, and (c) update properties of physical operator objects. We elaborate on them in turn.

Retrieval of specific properties of physical operators follows \textsc{sql}-like \textsf{SELECT-FROM-WHERE} syntax. The \textsf{SELECT-FROM} clauses are mandatory for any retrieval task. Predicates in the \textsf{WHERE} clause are formulated upon attributes of \textsc{poem} objects. Every query result is a set of \textsc{poem} objects with specific attributes specified in the \textsf{SELECT} clause and satisfies the conditions in the \textsf{WHERE} clause. The following example shows the retrieval of  the \textsf{ZZJOIN} operator object with \textsf{defn} attribute.

\begin{quote}
\begin{verbatim}
SELECT defn FROM pg WHERE name = 'zzjoin'
\end{verbatim}
\end{quote}

The following example shows retrieval of all objects representing the join operation in the source.
\begin{quote}
\begin{verbatim}
SELECT * FROM pg WHERE name LIKE '%join'
\end{verbatim}
\end{quote}
Our framework also supports join queries especially between physical objects from multiple \textsc{dbms}.

\textsc{pool} supports a \textsf{COMPOSE} clause to specify generation of a natural language description \textit{template} of an operator.
Specifically, the \textsf{COMPOSE} clause uses the \textsf{desc}, \textsf{type}, and \textsf{cond} attributes of operators to generate the template. For example, the template generation for the \textsf{HASH} operator can be specified as follows.
\begin{quote}
\begin{verbatim}
COMPOSE hash FROM pg
\end{verbatim}
\end{quote}

The above state will return the template \textit{``hash \$$R_1$\$''}, which can be subsequently used by our framework to generate specific description of the \textsf{HASH} operator in a \textsc{qep}. Note that the \textsf{COMPOSE} operator returns a value of type \textsf{string} instead of a \textsc{poem} object. Also, observe that $R_1$ is appended based on the \textsf{type} attribute of the \textsf{hash} object.

As mentioned above, \textsc{pool} allows composing a pair of critical and auxiliary operators (\eg hash and hash join) to generate the natural language description template for the critical operator.
\begin{quote}
\begin{verbatim}
COMPOSE hash, hashjoin FROM pg
USING hashjoin.desc = 'perform hash join'
\end{verbatim}
\end{quote}
The above statement generates the following template: \textit{``hash \$$R_1$\$ and perform hash join on \$$R_2$\$ and \$$R_1$\$ on condition \$$cond$\$''}. Since an operator object may have multiple \textsf{desc} attributes, the optional \textsf{USING} clause allows one to specify which one to use to generate the template. In the case it is unspecified, a \textsf{desc} will be chosen randomly for each operator in the \textsf{COMPOSE} clause to create the template. Hence, the form of a \textsf{COMPOSE} statement is: \textsf{COMPOSE} \textit{<list of object names>} \textsf{FROM} \textit{<source>} \textsf{USING} \textit{<condition on \textsf{desc}>}.  Note that in the case the list of object names contains more than one operators, it must be an (auxiliary, critical) operator pair, which generates the template for the critical operator.

\textsc{pool} supports update statements that allow attributes of existing \textsc{poem} objects to be changed. The general form of the update statement is: \textsf{UPDATE} \textit{<source>} \textsf{SET} \textit{<new-value assignments> }\textsf{WHERE} \textit{<condition>}. Each new-value assignment is an attribute, an equal sign and a string. If there are more than one assignment, they are separated by commas. The following example shows updating the definition of the hash join operator.

\begin{quote}
\begin{verbatim}
UPDATE pg
SET defn = 'a type of join algorithm...'
WHERE name = 'hashjoin'
\end{verbatim}
\end{quote}

The update statement can be exploited to assign definition or description of an operator from one commercial database to another, thereby making it more efficient for an \textsc{sme} to specify properties of physical operators. The following example demonstrates how description of hash join in PostgreSQL is transferred to the hash join operator in DB2.

\begin{quote}
\begin{verbatim}
UPDATE db2
SET desc = (SELECT desc
 FROM pg WHERE pg.name = 'hashjoin')
WHERE db2.name = 'hsjoin'
\end{verbatim}
\end{quote}

It can also be used along with the \textsf{REPLACE} clause to transfer definition or description of an operator to another within the \emph{same} source. For example, one can transfer the description of hash join to nested loop join by replacing the word \textsf{`hash'} with \textsf{`nested loop'} as follows.

\begin{quote}
\begin{verbatim}
UPDATE pg
SET desc = REPLACE((SELECT desc FROM pg AS pg2
WHERE pg2.name = 'hashjoin'), 'hash', 'nested loop')
WHERE pg.name = 'nested loop join'
\end{verbatim}
\end{quote}

Note that the \textsf{REPLACE} clause takes three parameters as input, namely the description or definition of a \textsc{poem} object, the string in it that needs to be replaced (\eg \textsf{`hash'}), and its new replacement string (\eg \textsf{`nested loop'}).

\vspace{1ex}\noindent\textbf{Implementation.}
\textsc{pool} is implemented on top of a standard relational database.  \textsc{poem} objects are stored in two relations with the following schema: \textsf{POperators(\underline{oid}, source, name, alias, type, defn, cond, targetid)} and \textsf{PDesc(\underline{oid, desc})} as an object may have multiple descriptions. We use Python script to translate \textsc{pool} statements to corresponding \textsc{sql} statements on these relations. A wrapper is implemented that takes the results of \textsc{sql} queries as input and returns \textsc{poem} objects or string as output.
\vspace{0ex}
\section{The Framework of Rule-LANTERN} \label{sec:rule}
Our rule-based framework, \textsc{rule-lantern}, leverages the narration (descriptions by \textsc{sme}s) of various operators defined using \textsc{pool} to generate a natural language description of the \textsc{qep} of an \textsc{sql} query. In this section, we describe it in detail.  We begin with the model of the framework for generating natural language descriptions of \textsc{qep}s. Next, we highlight the key issues for designing  \textsc{rule-lantern}. Finally, we present the algorithm for realizing \textsc{rule-lantern}.

\vspace{0ex}
\subsection{Model For Narration of QEPs} \label{sec:model}
Narration is the use of techniques to convey a story to an audience~\cite{narration}. In our context, the story is the description of a query execution plan and the audience consists of learners. Chatman~\cite{Chatman} defines \textit{narrative} as a \textit{story} (content of the narrative) and \textit{discourse} (expression of it). In a nutshell, the story can be viewed as the logical form of the narrative, while the discourse prunes out unimportant content and focuses on presenting components deemed interesting in a particular order. 

Inspired by this classical model of narration, El Outa \etal~\cite{OF+20}  recently proposed a four-layered model for data narration\footnote{\scriptsize The factual, intentional, structural, and the presentation layers map to \textit{form of content},  \textit{substance of content}, \textit{form of expression}, and \textit{substance of expression} of \cite{Chatman}, respectively.} that we adopt for \textsc{qep}s.
Specifically, the narration of \textsc{qep}s is modeled as follows.

\begin{itemize}
\item \textbf{Factual layer.} The factual layer models \textsc{qep}s (\ie data) using \textit{language-annotated operator trees} that allow for manipulation of \textsc{qep}s for narration generation.

\item\textbf{Intentional layer.} The intentional layer models the substance of a story by identifying the content (description of various operators) based on the desired goal (\ie comprehension of a \textsc{qep} by learners).

\item \textbf{Structural layer.} The structural layer models the structure of a narrative by organizing its \textit{plot} (\ie the arrangement of messages in a way easily understandable by the audience). While the previous layers focus on the contents of the narrative, this layer focuses on its discourse. In our framework, we organize the plot as a sequence of \textit{steps}.

\item \textbf{Presentation layer.} It models the presentation of a narrative. That is, how a story is presented to the audience.
\end{itemize}

Our \textsc{rule-lantern} addresses the first three layers. We utilize the presentation approach of~\cite{neuron} for the presentation layer.

\vspace{0ex}
\subsection{Design Issues} \label{sec:design}
 At first glance,  we may simply perform a post-order traversal of an operator tree and exploit the natural language description templates specified using \textsc{pool}  to transform the information contained in each node ``independently'' to its natural language description and simply aggregate them to generate the description of a \textsc{qep}. This method, however, may produce a verbose description containing redundant information. This is because a node in an operator tree may only convey a segment of  information related to a specific physical operator. For instance, in PostgreSQL, the node representing \textsf{HASH JOIN} have a child called \textsf{HASH}. The latter node conveys  the hashed relation and can be considered as a part of the main narrative of performing hash join between two relations. Hence, it is important to consider the roles played by different nodes for generating concise natural language descriptions of \textsc{qep}s.

A consequence of the above issue is that the natural language description of an execution step related to a specific operation may need to be generated by \textit{composing} descriptions  of multiple nodes. For example, consider the \textsf{TBSCAN} operator (\ie table scan operator in DB2). In one \textsc{qep}, we may simply perform a table scan on a relation without any filtering condition. On the other hand, in another \textsc{qep}, we may perform a table scan on a relation based on certain filtering condition (\eg \textsf{title contains `July'}) using the \textsf{FILTER} operator.  Hence, in the former plan, the natural language description is simply based on the  \textsf{TBSCAN} node whereas in the latter plan, concise description demands \textit{composition} of \textsf{ desc} attributes of \textsf{TBSCAN} and   \textsf{FILTER} nodes.  Hence, \textsc{rule-lantern} should support such composition.

\eat{Third, a \textsc{qep} may utilize the iterator, materialization, or vectorization model for query execution. For example, PostgreSQL use \textsf{materialize} node to specify that the results of its child should be stored in the memory. Hence, such information in a \textsc{qep} should be clearly conveyed to a learner. }

\vspace{0ex}
\subsection{Language-annotated Operator Tree}
Observe that an operator tree does not contain any information related to natural language descriptions of the operators. Hence, we extend it to annotate the nodes with their natural language descriptions as specified using \textsc{pool}. We refer to such extension as
\textit{language-annotated operator tree} (\textsc{lot}), denoted as $T_L = (N, E)$.  Formally, each node $n \in N$ in $T_L$ is associated with a\textit{ name}, denoted as $n.name$,  and a \textit{label}, denoted as $n.label$. The former  is set to the \textsf{alias} value of the corresponding object in \textsc{poem}. In the case the \textsf{alias} is unspecified, it is set to the object's \textsf{name}. The latter contains a natural language description of $n$ generated from the natural language template created by executing \textsf{COMPOSE} statement of \textsc{pool} on $n$. \eat{Note that by default $n.label =$  ``$n.name\ R$'' where $R$ is the relation on which the operation is performed.} For example, reconsider the operator tree in Figure~\ref{fig:exp-qep}. For the node representing hash join, $n.name = \textsf{HASH JOIN}$. A natural language description of this node can be $n.label =$ \textit{``perform hash join on table \$$R_1$\$ and table \$$R_2$\$ on condition \$$C$\$}'' where $R_1$ (resp. $R_2$) and $C$ are place holders for input relations and join condition(s), respectively. Note that this template is returned by the following \textsc{pool} query: \textsf{COMPOSE hashjoin FROM pg}. Also, specific relation/attribute names and conditions to replace the placeholders are added in subsequent steps.\eat{ Observe that this component realizes the factual layer of the model in Section~\ref{sec:model}. }\eat{ On the other hand, suppose the node \textsf{HASH} does not have any user-specified label. In this case, $n.label =$ ``hash  \$$R_1$\$'' . In the next section, we shall see how to declaratively specify the labels of these nodes. Note that the name of a physical operator may be different across different commercial databases. For instance, \textsf{HASH JOIN} operator in PostgreSQL is called \textsf{HSJOIN} in IBM DB2\footnote{\scriptsize \url{https://www.ibm.com/support/knowledgecenter/SSEPGG_11.5.0/com.ibm.db2.luw.admin.explain.doc/doc/r0052023.html}.}. Furthermore, some operators may exist in one commercial DBMS but not in another. For example, \textsf{FILTER} or \textsf{ZZJOIN} (zigzag join) operator exist in DB2 but not in PostgreSQL.}

\vspace{0ex}
\subsection{Composition of Node Labels} \label{sec:compose}
To tackle the issues described in Section~\ref{sec:design}, we logically \textit{refine} a \textsc{lot} by clustering the auxiliary nodes with the corresponding critical ones. Recall that these two types of nodes are specified by an \textsc{sme} using \textsc{pool}. For example, in PostgreSQL, \textsf{HASH JOIN} node and its child \textsf{HASH}, \textsf{MERGE JOIN} node and its child \textsf{SORT} are two examples of auxiliary-critical node pairs that can be specified using \textsc{pool}.

 Given a \textsc{lot} $T_N = (N, E)$, $cluster(T_N)$ returns a set pairs of nodes in $T_N$, $\{(n_a, n_c)| (n_a, n_c) \in E \wedge n_a \neq n_c\}$, where $n_c$ and $n_a$ denote critical and auxiliary nodes, respectively.  Each pair of critical and auxiliary nodes in a cluster is translated into a\textit{ single} natural language description template using the \textsf{COMPOSE} statement as an auxiliary node contributes to a segment in the description. For example, consider the \textsf{HASH JOIN} and its child \textsf{HASH} in a \textsc{lot}.  A natural language description template of this pair of nodes can be as follows: \textit{``hash \$$R_1$\$ and perform hash join on \$$R_2$\$ and \$$R_1$\$ on condition \$$cond$\$''}. Observe that the segment \textit{``hash \$$R_1$\$''} is generated from the \textsf{HASH} node.

Under the hood, the \textsf{COMPOSE} statement on a pair of nodes is realized using a \textit{composition} operator, denoted by $\circ$. Given a pair of critical and auxiliary nodes $(n_a, n_c)$ such that $(n_a, n_c) \in E$, $n_a \circ n_c = n_a.label \wedge n_c.label$ where $\wedge$ represents \textit{``and'}'. In the above example, $n_a.label =$ \textit{``hash \$$R_1$\$''} and $n_c.label =$ \textit{``perform hash join on \$$R_2$\$ and \$$R_1$\$ on condition \$$cond$\$''}. The composition operator is neither associative nor commutative. The left operand must be an auxiliary node. This is intuitive as the natural language description is unclear if  \textit{``hash \$$R_1$\$''} appears after the label of \textsf{HASH JOIN}.

\vspace{0ex}
\subsection{Algorithm}

Algorithm~\ref{rule_based} outlines the procedure for generating a natural language description of a \textsc{qep} in \textsc{rule-lantern}. It first extends the operator tree $T$ to a \textsc{lot} $T_L$ (Line 1). Observe that in a graphical representation of a \textsc{qep} (\eg Figure~\ref{fig:plan2}), hierarchical relations between operators and data flow are clearly indicated by edges in the tree. In contrast, a natural language description is inherently sequential as a reader reads it top-down like a document. Particularly, a parent of an operator may not be translated immediately as the next step during natural language generation  as other children need to be translated first (\eg auxiliary nodes). Therefore, in order to ensure clarity of data flow, this step also assigns a unique \textit{identifier} to the output of each operator (\ie intermediate results) so that it can be appropriately referred to in the translation of its parent (denoted by $node.identifier$). For example, in Figure~\ref{fig:exp-qep}, the intermediate results  of the \textsf{SEQSCAN} operation on the \textsf{Publication} relation is assigned an identifier $T_1$. This identifier  is subsequently used in the natural language description of its parent \textsf{HASH} node.

Next, it retrieves the cluster $C$ in $T_L$ containing a set of critical and auxiliary node pairs by leveraging the \textsc{poem} store (Line 2). Since the structural layer of our model consists of a sequence of steps to describe the \textsc{qep}, it traverses $T_L$ in post-order manner to generate these steps. If a node and its child are an element in $C$ then it uses the label of the critical node to generate corresponding natural language description by replacing the place holders with corresponding values (Lines 5-6).\eat{ Note that if a node $n$ is an auxiliary node and does not have an user-specified label, then its description is generated from $n.name$.}  For example, in Figure~\ref{fig:exp-qep},  the \textsf{HASHJOIN} node satisfies the condition in Line 5 and is translated as follows: \textit{``hash $T_1$ and perform hash join on  \textsf{inproceedings} and $ T_1$ on condition (\textsf{(i.proceeding\_key)= (p.pub\_key)})''}. Observe that $T_1$ is the identifier of intermediate results of the \textsf{SEQSCAN} node. On the other hand, if a node is not in $C$, then the corresponding $step$ is generated by utilizing  its label.  For example, the right leaf node is translated to\textit{ ``perform sequential scan on \textsf{inproceedings}''}.   Finally, the intermediate relation information  is appended to $step$ in Lines 10-11. For instance, the segment \textit{``to get the intermediate relation $T_2$''} is appended to the above $step$ of the \textsf{HASHJOIN} node. In the case, the node represents the final operation in an operator tree, \textit{``to get the final results."} is appended to $step$ (Line 13). Observe that contents of these nodes represent the intentional layer of our model. The time complexity of generating a natural language description is $O(N)$.

\begin{algorithm}[t]   \small
	\caption{\textsc{rule-lantern} Algorithm}
	\label{rule_based}
	{\bf Input:	An operator tree $T = (N, E)$, \textsc{poem} store $P$;}\\
	 {\bf Output:	Natural language translation $result$;}\\
	\begin{algorithmic}[1] \small
	\STATE $T_L \leftarrow$ \textsc{GenerateLOT($T, P$)}
	\STATE $C \leftarrow$ \textsc{Cluster($T_L$, $P$)}
		 \FORALL{$node\in T_L$}
		 	\STATE $step \leftarrow  \emptyset$
		 	\IF {$(node.child, node) \in C$}
		 	    \STATE $step \leftarrow$ \textsc{Translate($node.child, node, step$)}
		 	  \ELSE
                \STATE $step \leftarrow$  \textsc{Translate($node.label, step$)}
            \ENDIF
		 	\IF {$node.parent  != \emptyset$ and $node.identifier != \emptyset$}
		 	    \STATE $step \leftarrow$ \textsc{AppendIntermediate($step$, $node.identifier$)}
		 	  \ELSIF{$node.parent  = \emptyset$}
		 	  \STATE $step \leftarrow$ \textsc{AppendFinal}($step$, \textsf{``to get the final results.'}' )
		 	   \ENDIF
		 	\STATE $result \leftarrow$ \textsc{Add}($result$, $step$)
		 \ENDFOR
		 \RETURN $result$		
	\end{algorithmic}
\end{algorithm}

\eat{\begin{itemize}
    \item The column(s) used in \textsf{SORT}, \textsf{AGGREGATE}, and \textsf{HASH}.
    \item The join condition used by any \textsf{JOIN}.
    \item The filtering condition used by any \textsf{SCAN} and \textsf{AGGREGATE}.
    \item Whether an operator is run in parallel.
\end{itemize}}

\begin{example} \label{eg:2}
Consider the operator tree in Example~\ref{eg:1}. The \textsc{rule-lantern} algorithm generates the description of the \textsc{qep} as the following sequence of steps. (1) Visit \textsf{SEQ SCAN} for table \textsf{inproceedings} and generate\textit{ ``perform sequential scan on \textsf{inproceedings}."} (Line 8). Note that the $identifier$ of intermediate results associated with this node is set to \textsf{null} as there is no  filtering condition (\ie intermediate relation is identical to the base relation).  (2) Visit \textsf{SEQ SCAN} for table \textsf{publication} and generate \textit{``perform sequential scan on \textsf{publication} and filtering on \textsf{(title  containing 'July')} to get the intermediate relation $T_1$."} (Lines 8, 11). (3) Visit \textsf{HASH JOIN} and generate \textit{``hash table $T_1$ and perform hash join on \textsf{inproceedings} and $T_1$ on condition \textsf{((i.proceeding\_key)= (p.pub\_key))} to get the intermediate relation $T_2$."} (Lines 6, 11). (4) Visit \textsf{AGGREGATE} and generate \textit{``sort $T_2$ and perform aggregate on $T_2$ with grouping on attribute \textsf{i.proceeding\_key} and filtering on \textsf{(count(all) > 200)} to get the intermediate relation $T_3$."} (Lines 6, 11). (5) Visit \textsf{UNIQUE} and generate \textit{``perform duplicate removal on $ T_3$ to get the final results."} (Lines 8, 13).
\EndOfProof\end{example}

\textbf{Remark.} It is worth noting that the aforementioned algorithm is generic and can be realized on any commercial \textsc{rdbms}. Specifically, although the physical operator names are different across different \textsc{rdbms}, the \textsc{rule-lantern} framework operates on  $T_L$ and $C$, which are generated using the \textsc{poem} store.\eat{  Also, the textual description of a \textsc{qep} is richer in implementation-specific information of a query compared to textual narrative generated from a declarative \textsc{sql} query by tools like Logos \cite{KV+12}. This is because execution-specific details (\eg type of join, type of scan) of an \textsc{sql} query cannot be simply gleaned from its declarative statement.}\eat{ Note that in the presentation layer, we can hyperlink each operator name in a description with its corresponding definition retrieved by issuing \textsc{pool} queries. This further enables a learner to understand an unfamiliar operator.}

\vspace{0ex}
\section{Neural-LANTERN Framework}\label{sec:neural}
Although the \textsc{rule-lantern} technique can generate accurate natural language descriptions of \textsc{qep}s, our engagement with learners revealed an intriguing problem of this approach. Since the natural language description of an operation is generated from \textsc{sme}-specified descriptions in the \textsc{poem} store, the descriptions in \textsc{qep}s can be repetitive and lack variability. For example, the description in Step 3 of Example~\ref{eg:2} will be repeated for all \textsc{qep}s containing a hash join operator although the input relations or join conditions may differ. Note that even though \textsc{pool} allows an \textsc{sme} to specify multiple descriptions of an operator, in practice she may only specify one.  Consequently, some learners found that after reading the descriptions for several queries, they feel bored due to the usage of the same language to describe the operations.  They reported that they started skipping  several sentences in the descriptions. In fact, this is consistent with research in psychology that have found that repetition of text messages can lead to annoyance and boredom~\cite{CP79} resulting in purposeful avoidance~\cite{HK13}, content blindness~\cite{HG+11}, and even lower motivation~\cite{SPC90}.\eat{ In addition, from the development perspective, the rule-based generation requires human expertise for every domain of interests, which may be prohibitively expensive and not scalable to many domains.} To mitigate this problem, we propose a novel neural network-based framework called \textsc{neural-lantern} that is inspired by theories from psychology.

\vspace{0ex}
\subsection{Habituation and Boredom} \label{sec:psycho}
In psychology theory, \textit{habituation} is a decrease in response to a stimulus after repeated presentations~\cite{habituation}. The advantage of habituation is that it enables individuals to tune out unimportant information to be more productive or efficient. However, it also creates boredom that makes an individual disinterested in the information. Specifically, many studies in psychology such as ~\cite{ohanlon81} reported that habituation of cortical arousal in response to repetitive stimulation contribute to the likelihood that boredom\footnote{\scriptsize Mikulas and Vodanovich~\cite{MV93} defined boredom as ``a state of relatively low arousal and dissatisfaction, which is attributed to an inadequately stimulating situation''. Watt and Vodanovich~\cite{WV99} describe boredom as a dislike of repetition or of routine.} is experienced. In fact, subjectively monotonous activities could lead to a high degree of frustration and boredom~\cite{HP85}. Simple and homogeneous stimulus (e.g., same or similar messages) as well as high
exposure, accelerate the appearance of boredom~\cite{HC72}.

Diverse messaging has been studied to mitigate the problems germinated from repeated exposure. In controlled experiments, diversification was shown to reduce tedium from repeated exposure~\cite{HC72,SPC90}.  However, the messages were manually developed in these studies. Recall that \textsc{pool} also allows specifying such manual description using multiple \textsf{desc} attribute values. However, such manual generation creates a major barrier for diversifying descriptions of operators in \textsc{qep}s. Hence, systematic and automated technique is necessary for mitigating the negative effects of repeated exposure of similar descriptions.
\vspace{0ex}
\subsection{Training Data Generation}\label{specialenvir}
If we regard a \textsc{qep} as an input language while the natural language description as the output, interpreting \textsc{qep} into natural language can be viewed as a machine translation task, which can be addressed by a deep learning-based framework. However, as remarked earlier, it brings in two key challenges. First, it is prohibitively expensive to get large volumes of training data for this task. Second, the description generated as output should mitigate the appearance of boredom among learners when they peruse the natural language descriptions of \textsc{qep}s. We address these challenges by presenting a neural network-based framework called
\textsc{neural-lantern}.\eat{ Specifically, it leverage on \textit{Seq2Seq} to translate a given \textsc{qep} into its natural language description.} We begin with the training data generation process in this framework.

The training sample of a translation task consists of two parts, the sentence in the original (\resp input) language to be translated, \ie the input operator tree, as well as the ground-truth translation in the output language, \ie the natural language description of the corresponding \textsc{qep}. We shall now discuss these parts in turn.

\vspace{1ex} \noindent\textbf{Input.}  Given a relational database, we need a large number of \textsc{sql} statements in order to generate a large number of corresponding \textsc{qep}s for translation. Hence, we adopt the approach in~\cite{DBLP:conf/cidr/KipfKRLBK19} to generate a set of \textsc{sql} queries given a particular schema and database instance. This enables us to generate thousands of queries given a relational database instance. These queries contain aggregation, projection, as well as various filtering and join predicates. 

\eat{\begin{figure}[t]
		\centering
		\includegraphics[width=\linewidth]{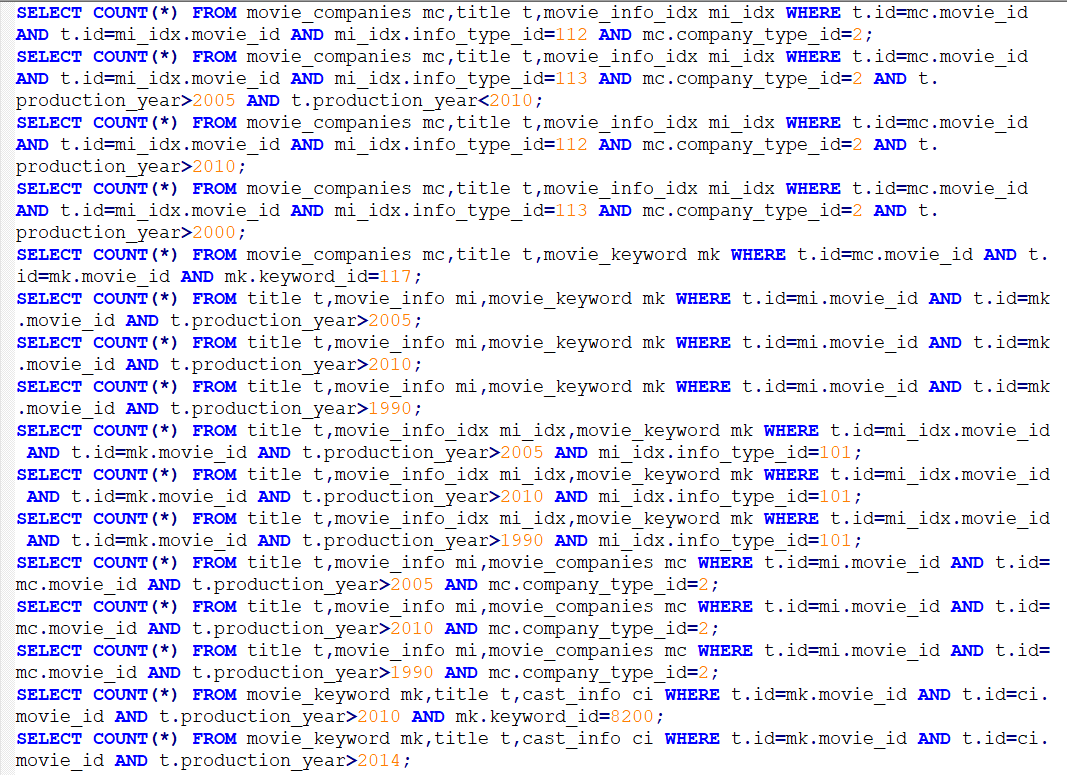}
		\vspace{-1ex}\caption{Example of the generated SQL statements.}
		\label{fig:sqlgeneg}
		\vspace{-2ex}\end{figure}}

Next, we  acquire a collection of \textsc{qep}s corresponding to these queries. Each \textsc{qep} is decomposed into a set of \textit{acts} (denoted as $actCol$), each of which corresponds to a set of operators over some relations. For instance, in Figure~\ref{fig:exp-qep}, \textsf{SEQUENTIAL SCAN}  and  (\textsf{HASH JOIN}, \textsf{HASH}) are two acts.  Specifically, each act is a single node (\ie operator) or a cluster (recall from Section~\ref{sec:compose}) in an operator tree. Each of such act, in the form of some operators and corresponding relations/conditions, is employed as an input training sample, and its corresponding natural language description is used as output sequence in the translation model. Specifically, for each act, we employ the strategy in \textsc{rule-lantern} to generate the corresponding \textsc{lot} in order to generate its corresponding natural language description. Observe that our input is at the act-level (\ie a subtree of an operator tree) instead of the entire operator tree. This enables us to not only generate numerous training data at specific operator-level but also, as we shall see in the next subsection, facilitates injecting diversity in the natural language description of each act, which in turn improves the neural model generalization.

\vspace{1ex} \noindent\textbf{Output.} For each training act, we have to obtain its natural language translation as the output sequence. We apply \textsc{rule-lantern} to generate the natural language description for each input. Notably, we need to pay attention to the schema-dependent variables, \eg relation/column names and filtering conditions, which do not contribute to the training of a translation model. We mark them with special symbols in the output labels for each input operation.  For example, an \textsf{INDEX SCAN} node is translated by \textsc{rule-lantern} into the followings: ``\textit{perform index scan on \$$R_1$\$ and filtering on \$$C$\$ to get the intermediate relation \$$R_2$\$}''. We replace it with ``\textit{perform index scan on $<$T$>$ and filtering on $<$F$>$ to get the intermediate relation $<$TN$>$}'' in the output label. Herein, special tags are adopted in the output to replace specific relations or predicates. The set of special tags we use is shown in Table \ref{tb:mark_symbol}. This leads us to a set of training samples, each of which consists of an operation (\ie act) as input and a corresponding \textsc{nl} description as output.\eat{ Examples for variate operations and their corresponding output in the training samples are illustrated in Table~\ref{tb:input_and_output_data}.}

\begin{table}[t] \scriptsize
	\caption{List of special tags used in the output.}
	\vspace{-3ex}\begin{center}
		\begin{tabular}{|l|p{3cm}|p{3cm}|}
			\hline
			\textbf{Tag} & \textbf{Description} & \textbf{Example}\\
			\hline
			$<$I$>$ & indexed column name &  \\
			\hline
			$<$F$>$ & filtering condition & $c\_mktsegment$ = 'BUILDING'\\
			\hline
			$<$C$>$ & join condition & $c\_custkey = o\_custkey$ \\
			\hline
			$<$T$>$ & an existing temporary table name &  \\
			\hline
			$<$TN$>$ & new temporary table name &  \\
			\hline
			$<$A$>$ & column name for sort & order by \textit{revenue} desc ...\\
			\hline
			$<$G$>$ & column name for groupby &group by $l\_orderkey$ ... \\
			\hline
		\end{tabular}
		\label{tb:mark_symbol}
	\end{center}
\vspace{0ex}\end{table}

\vspace{0ex}
\subsection{Diversifying Translation}\label{ssec:diversify}
The preceding subsection describes a strategy to generate training data automatically instead of manual labeling. Specifically, the output labels for the training samples are all generated by \textsc{rule-lantern}. Consequently, it does not address two key challenges discussed earlier.  First, the translation generated by \textsc{rule-lantern} can be repetitive leading to possible boredom among learners while reading the natural language descriptions of \textsc{qep}s.  Second, the amount of training data generated is still limited since \textsc{rule-lantern} imposes a one-to-one mapping between an act and its corresponding natural language description.\eat{ Note that training deep models with limited data is likely to suffer from overfitting.}

To address these challenges, we employ three popular state-of-the-art synonymous sentence generation tools~\cite{synonymous1,synonymous2,synonymous3} to expand the training samples as well as inject diversity in the translated text.  For the same \textsc{sql} statement, these models can generate a variety of natural language descriptions that add diversity to the narrative. In particular, for each of the \textsc{rule-lantern} results, we apply all the three tools and acquire a set of synonymous sentences. Notably, we remove duplicates (if any) and manually eliminate invalid sentences  generated by these tools. 
As a result, we enlarge the number of training samples in our datasets by approximately 3 times. 

Table \ref{result} shows an example of three synonymous sentences  generated by these tools from a \textsc{rule-lantern}-generated text.
An interesting observation is that these tools may not always choose correct words in the generated sentences. For example, in sentences 1 and 2, the word ``separating'' is generated instead of ``selecting''.  At first glance, it may seem that this may hinder a learner's comprehension. However, surprisingly, our empirical study shows that is not the case. Instead, it may even arouse interest among learners as they encounter novel unexpected words\eat{ that they do not expect to encounter in the text}.

\vspace{0ex}
\subsection{Translation Model}

\textbf{Task definition.} To finish our translation task, we present a \textit{QEP2Seq} model following the Seq2Seq structure. For a  \textsc{qep}, the \textit{acts} collection $actCol$ is composed of a series of acts ${L_1, L_2, \ldots, L_n}$, each of which is derived from the \textsc{qep}. Specifically, each act $L_i$ constitutes an input to the neural Seq2Seq model, and the corresponding output is the generated description $S_i$ containing $m$ tokens ${o_1, o_2, \ldots, o_m}$, with $o_t$ being the word at position $t$. Our goal is to train a model parameterized by $\theta$ that can be used to infer the most likely natural language description $S_i$ for any given input $L_i$ as follows:
\vspace{-1ex}\begin{equation}
	\label{pro_eq} 
	\hat{S}_i = \mathop{\arg\max}_{y_{{1:m}}} \prod_{t=1}^{m} P_{\theta} (y_t = o_t|y_{0:t-1}, L_{i})
\end{equation}
{The explanation for the entire \textsc{qep} containing acts ${L_1, L_2, \ldots, L_n}$  is then constructed by concatenating the $\hat{S}_i$'s for $i = 1 \ldots n$.}

\begin{table}[t] \scriptsize
	\caption{Examples of synonymous sentence generation.}
	\vspace{-3ex}\begin{tabular}{|p{2cm}|p{6cm}|}  
		\hline  
		\textbf{Approach} & \textbf{Description} \\ \hline  
		\textsc{rule-lantern} & perform sequential scan on user and filtering on age $>$ 10 to get the final results. \\ \hline  
		synonymous sentence 1 &  perform sequential scan on user and separating on age $>$ 10 to get the conclusive outcome. \\  
		\hline  
		synonymous sentence 2 &  execute sequential scan on user and separating on age $>$ 10 to get the conclusive outcome. \\  
		\hline  
		synonymous sentence 3 &  execute sequential scan output on user and get user which age $>$ 10 and to get the conclusive outcome. \\  
		\hline  
	\end{tabular}\label{result}
\end{table}




\vspace{-1ex}\subsubsection{Model Architecture} 

As shown in Figure~\ref{model}, our QEP2Seq model consists of an \emph{Encoder} and a \emph{Decoder}. The decoder employs an attention mechanism so that it can focus on the relevant portion of the input when generating a target word.  Besides, we also use pre-trained word vectors in the \textit{Decoder} (see \emph{Embedding} in Figure~\ref{model}).

\vspace{1ex} \noindent\textbf{Pre-trained word vectors.} Static and contextualized pre-trained word representations like GloVe \cite{pennington2014glove}, Word2Vec \cite{mikolov2013efficient} and BERT \cite{devlin2018bert} have attracted a great amount of attention recently in \textsc{nlp}. The vector representations of words learned by these models have been shown to carry semantic information that can help the model to generalize well for different \textsc{nlp} tasks. 

In this work, we adopt both static (\emph{Word2Vec} and \emph{GloVe}) and contextual word embeddings (ELMo \cite{elmo} and BERT). While  static word embeddings are easy to use, using contextual embeddings effectively can be non-trivial. For ELMo, we take the embeddings from its two bi-\textsc{lstm} layers (each of size 4096) and take a linear combination of the vectors as the pre-trained representation of a word. For BERT, we take the representation from its last layer. We use the BERT-based model, which has 12 layers with 768 hidden units and 12 heads. Empirical study in the next section demonstrates that using pre-trained word embeddings can accelerate the convergence of our QEP2Seq model and alleviate overfitting problem.

\vspace{1ex} \noindent\textbf{Encoder.} 
The \textit{Encoder} \textsc{rnn} encodes each word $w_t$ in $L_i$ into the corresponding hidden state $\mathbf{h}_t$ using an \textsc{lstm} layer. The \textsc{lstm} maintains a vector of \emph{memory cells} $\mathbf{c}_t \in \mathbb{R}^d$ (a.k.a. \emph{cell state}) to store \emph{long term} memory, and uses \emph{(soft) gates} to control how much information to update with, to retrieve, or to remember for the next token. At each time step $t$, the \textsc{lstm} hidden layer receives previous hidden state $\mathbf{h}_{t-1}$ and the current input $\mathbf{x}_{t}$, i.e., the word embedding for token $w_t$ (randomly initialized). The $LSTM_{\text{Enc}} (\mathbf{h}_{t-1}, \mathbf{x}_{t})$ architecture used here is given by the following equations~\cite{graves2013speech}:
\begin{eqnarray}
	\mathbf{i}_t &=& \mathrm{sigmoid}(U_i\mathbf{h}_{t-1}^{l} + V_i\mathbf{x}_t)  \hspace{4em} \text{[input gate]} \label {lstm_first}\\
	\mathbf{f}_t &=& \mathrm{sigmoid}(U_f\mathbf{h}_{t-1}^{l} + V_f\mathbf{x}_t) \hspace{3.5em} \text{[forget gate]} \\
	\mathbf{o}_t &=& \mathrm{sigmoid}(U_o\mathbf{h}_{t-1}^{l} + V_o\mathbf{x}_t) \hspace{3.5em} \text{[output gate]}\\
	\mathbf{c}_t &=& \hspace{-0.1cm}  \mathbf{i}_t\odot \tanh(U_c\mathbf{h}_{t-1}^{l} + V_c\mathbf{x}_t) + \mathbf{f}_t\odot\mathbf{c}_{t-1} \hspace{0.5em} \text{[cell state]}\\
	\mathbf{h}_t &=& \mathbf{o}_t \odot \tanh(\mathbf{c}_t) \hspace{10em} \text{[output]} \label {lstm_last}
\end{eqnarray}

\begin{figure}[t]
	\centerline{\includegraphics[height=3.5cm, width = 0.9\columnwidth]{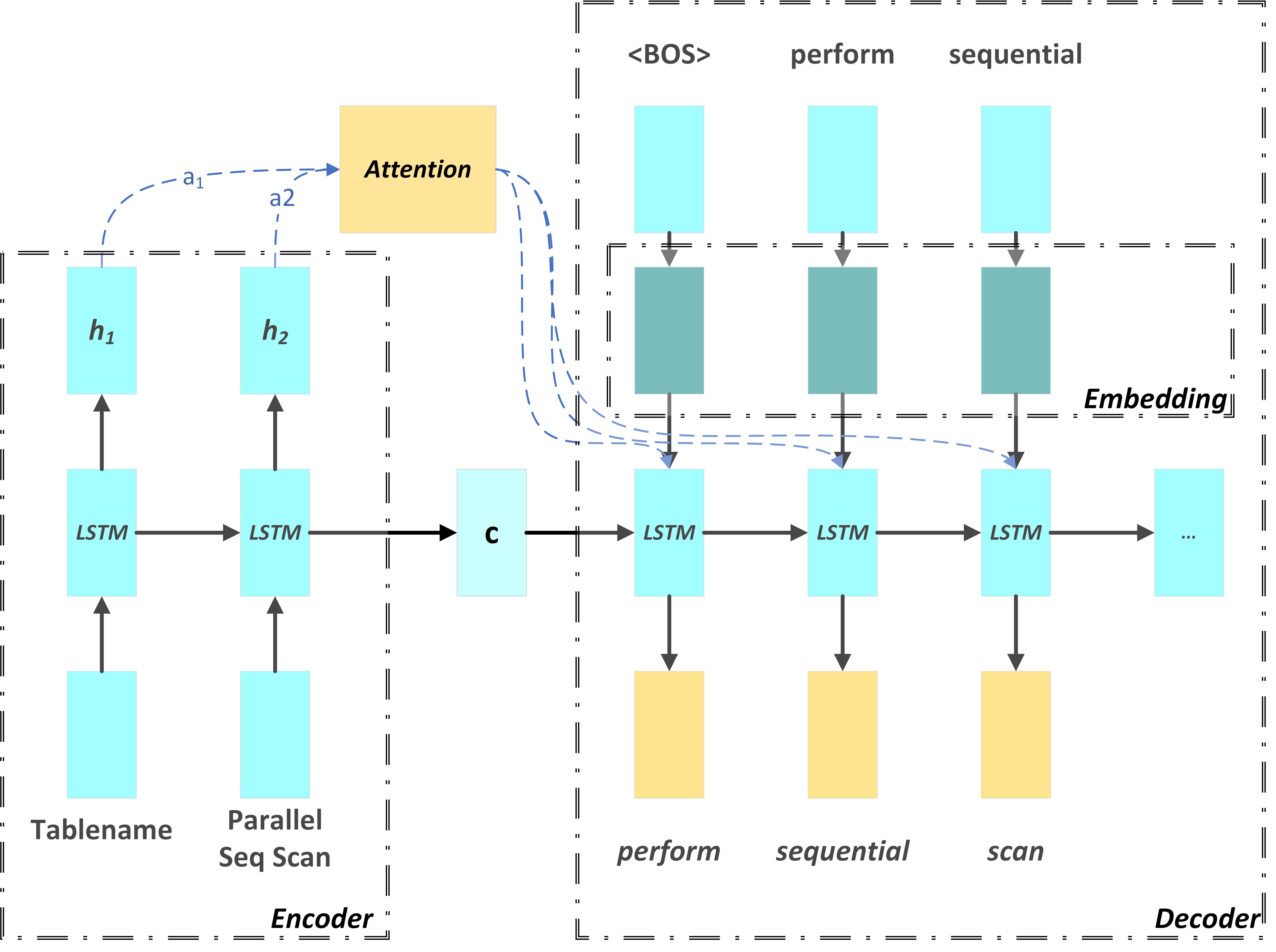}}
	\vspace{-2ex}\caption{The QEP2Seq Model.}
	\label{model}
\end{figure}

\noindent where  $U_{\cdot}$ and $V_{\cdot}$ are the corresponding weight matrices and $\odot$ denotes element-wise product. 

\vspace{1ex} \noindent\textbf{Decoder with attention.} To generate the natural language description ($S_i$) for the input sequence ($L_i$), we use an \textsc{lstm} decoder with an \emph{attention} mechanism \cite{luong2015effective}. The attention is an elegant way to let the decoder focus on the relevant portion of the encoder while generating a token (Figure~\ref{model}). Similar to the \textit{Encoder} \textsc{rnn}, at each time step $t$, the decoder \textsc{rnn} receives two inputs -- the previous hidden state $\mathbf{s}_{t-1}$ and the word embedding $\mathbf{x}_t$ (\eg Word2Vec, BERT). The \textsc{lstm} layer then constructs $\mathbf{s}_t$ following Equations \ref{lstm_first} - \ref{lstm_last} (using a different set of weight matrices).

\vspace{-1ex}\begin{equation}
	\mathbf{s}_t = LSTM_{\text{Dec}}(\mathbf{s}_{t-1}, \mathbf{x}_t)
\end{equation}

The decoder hidden state $\mathbf{s}_t$  is then used to compute a relevance score (attention weight) with respect to each of the encoder states $\mathbf{h}_t \textsf{ for } t = 1 \ldots N$ with $N$ being the number of tokens in $L_i$ (Eq. \ref{attentionSoftmax}).

\vspace{-1ex}\begin{equation}
	\label{attentionSoftmax}
	\alpha_{i}^t = \frac{\exp ( g (\mathbf{s}_t, \mathbf{h}_i) )} {\sum_{j=1}^{N} \exp (g(\mathbf{s}_t, \mathbf{h}_j))}
\end{equation}

In particular, $g (\mathbf{s}_t, \mathbf{h}_i)$ is a relevant score between the hidden state $\mathbf{s}_t$ of the \textit{Decoder} and the hidden state $\mathbf{h}_i$ of the \textit{Encoder}. There are several ways to compute the relevant scores. In our work, we use the \emph{additive} attention \cite{Bahdanau-attn} to measure the relevance score:

\vspace{-1ex}\begin{equation}
	g (\mathbf{s}_t, \mathbf{h}_i) = V_a^T \tanh (W_s \mathbf{s}_{t} + W_h \mathbf{h}_t)
\end{equation}

\noindent where $V_a, W_s$, and $W_h$ are learnable parameters. The attention weights are then used to compute a context vector $\mathbf{a}_{t}$ as a  weighted sum of encoder hidden states (Eq.~\ref{computeAttention}).
\vspace{-1ex}\begin{equation}
	\label{computeAttention}
	\mathbf{a}_{t} = \sum_{i=1}^{N}\alpha_{i}^t \mathbf{h}_i
\end{equation}

We concatenate the \textsc{lstm} state $\mathbf{h}_t$ and the context vector $\mathbf{a}_{t}$ and use the concatenated vector to compute the generation probability over the vocabulary items $o \in \mathcal{O}$ (Eq. \ref{finalsoft}).

\vspace{-1ex}\begin{equation}
	\label{finalsoft}
	P_{\theta} (y_t = o | y_{0:t-1}, L_{i} ) = \frac {\exp (\mathbf{w}_o^{T} [\mathbf{s}_{t};\mathbf{a}_{t}])}{ \sum_{o' \in \mathcal{O}} \exp (\mathbf{w}_{o'}^T [\mathbf{s}_{t};\mathbf{a}_{t}])}
\end{equation}

\noindent where $\mathbf{w}_o$ is the weight vector corresponding to the output word $o$.

\subsubsection{Model Training}

We minimize the cross entropy loss and use \emph{Teacher Forcing} \cite{williams1989teacherforcing} to train the Seq2Seq model. Teacher forcing, where current step's target token is passed as the next input to the decoder rather than the predicted token, is a common way to train neural text generation models for faster convergence. The loss for one input-output pair $(L_i, S_i)$  can be written as:
\begin{eqnarray}
	\mathcal{L}(\theta) = - \sum_{t=1}^{m} \sum_{o \in \mathcal{O}} \mathbbm{1}(y_t = o) \log P_{\theta} (y_t = o | y_{0:t-1}, L_{i} ) \label{logloss}
\end{eqnarray}

\noindent where $\mathbbm{1} (y_t=o)$ is an indicator function that returns $1$ if $y_t=o$ otherwise $0$. Our \emph{LSTM} layer has 256 cells at each layer, with an input vocabulary of 36 and an output vocabulary of 62. The statistics about the word embeddings and the resulting \emph{LSTM} parameters are listed in Table~\ref{tb:parasta}. The complete training details are given below:

\begin{table}[t]
	\begin{center} \scriptsize
		\begin{tabular}{|l|p{12mm}|l|p{25mm}|}
			\hline
			\textbf{Method} & \textbf{Dimension of embedding} & \textbf{\#parameters (total)} &\textbf{\#pure recurrent connections (Encoder, Decoder)} \\
			\hline
			\textit{QEP2Seq}+\emph{Word2Vec}	& 128 & 920,393 & 837,632 (279,552, 558,080)\\
			\hline
			\textit{QEP2Seq}+\emph{GloVe} & 100 & 993,901 & 907,264 (279,552, 627,712)\\
			\hline
			\textit{QEP2Seq}+\emph{BERT}& 768&1,716,009&1,591,296 (279,552, 1,311,744)  \\
			\hline
			\textit{QEP2Seq}+\emph{ELMo}&1024&1,992,745& 1,853,440 (279,552, 1,573,888)  \\
			\hline	
		\end{tabular}
		\vspace{0ex} \caption{Statistics about our \emph{LSTM} layer.}\label{tb:parasta}
	\end{center}
	\vspace{-1ex}\end{table}


\begin{itemize}
	\item We initialized all of the \textsc{lstm}'s parameters with the uniform distribution between -0.1 and 0.1
	\item We used stochastic gradient descent (\textsc{sgd}) without momentum, with a fixed learning rate of 0.001.  We trained our models for a total of 50 epochs.
	\item We used minibatches of 4 sequences. 
	\item The dimension of the word embedding in the \emph{Encoder} is 16, and at the \emph{Decoder} is 32 when no pre-trained word vector is employed (\ie for random initialization).
	\item We select our model based on the validation loss.
\end{itemize}

\subsubsection{Natural Language Generation} \label{Natural_Language_Generation}

After the QEP2Seq model is trained, the most likely description can be inferred (or decoded) by: 
\begin{equation}
	\label{pro_eq}
	\hat{S}_i = \mathop{\arg\max}_{y_{{1:m}}} \prod_{t=1}^{m} P_{\theta} (y_t|y_{0:t-1}, L_{i})
\end{equation}

\noindent  In the above equation, $\theta$ denotes the trained QEP2Seq model, and $ \prod_{t=1}^{m} P_{\theta} (y_t|y_{0:t-1}, L_{i})$ is the probability that the model assigns to sequence $y_{{1:m}}$ for an input sequence $L_i$. In practice, the \emph{argmax} procedure is intractable for large output vocabulary. To overcome that, we employ a Beam Search algorithm, which maintains a beam of $K$ partial hypothesis starting with the start symbol $<$BOS$>$ (as shown in Figure~\ref{model}). At each step, the beam is extended by one additional character and only the top $K$ hypotheses are kept. Decoding continues until the stop symbol $<$END$>$ is emitted, at which point the hypothesis is added to the set of completed hypotheses.

Finally, we replace the special tags (\eg $<I>,<C>,\ldots$) listed in Table~\ref{tb:mark_symbol} in the generated natural language using the corresponding identifiers.

\vspace{1ex} \noindent\textbf{Remark.} The \textsc{neural-lantern} framework is novel in the following ways. First, this is a seminal effort to model the \textsc{qep} to natural language description as a machine translation task. Second, our training data generation process is designed to address the psychological impact of repeated text on learners. Third, we propose a novel QEP2Seq scheme. 	As a \textsc{qep} cannot be regarded as an input sequence, we present a model to interpret \textsc{qep}s into a set of acts, each of which is viewed as an input for the translation model.\eat{	Besides, an attention layer and pre-trained static and contextualized word vectors are employed in the translation model, which further improve the effectiveness and generalization of the model.} 

\vspace{0ex}
\section{Experimental Study} \label{sec:perf}
\textsc{lantern} is implemented in Python.  In this section, we report the  performance results of \textsc{lantern}. All experiments are performed  on a server running Ubuntu 16.04.6 LTS with 2*Intel Xeon CPU E5-2680 v2 @ 2.80GHz, 256GB RAM, and 2*NVIDIA RTX 2080 Ti graphical card with 11GB GDDR6.

\begin{figure}[t]
	\centering
	\subfigure[Diversification of text]{
		\label{data_augmentation}
		\includegraphics[width=0.7\columnwidth]{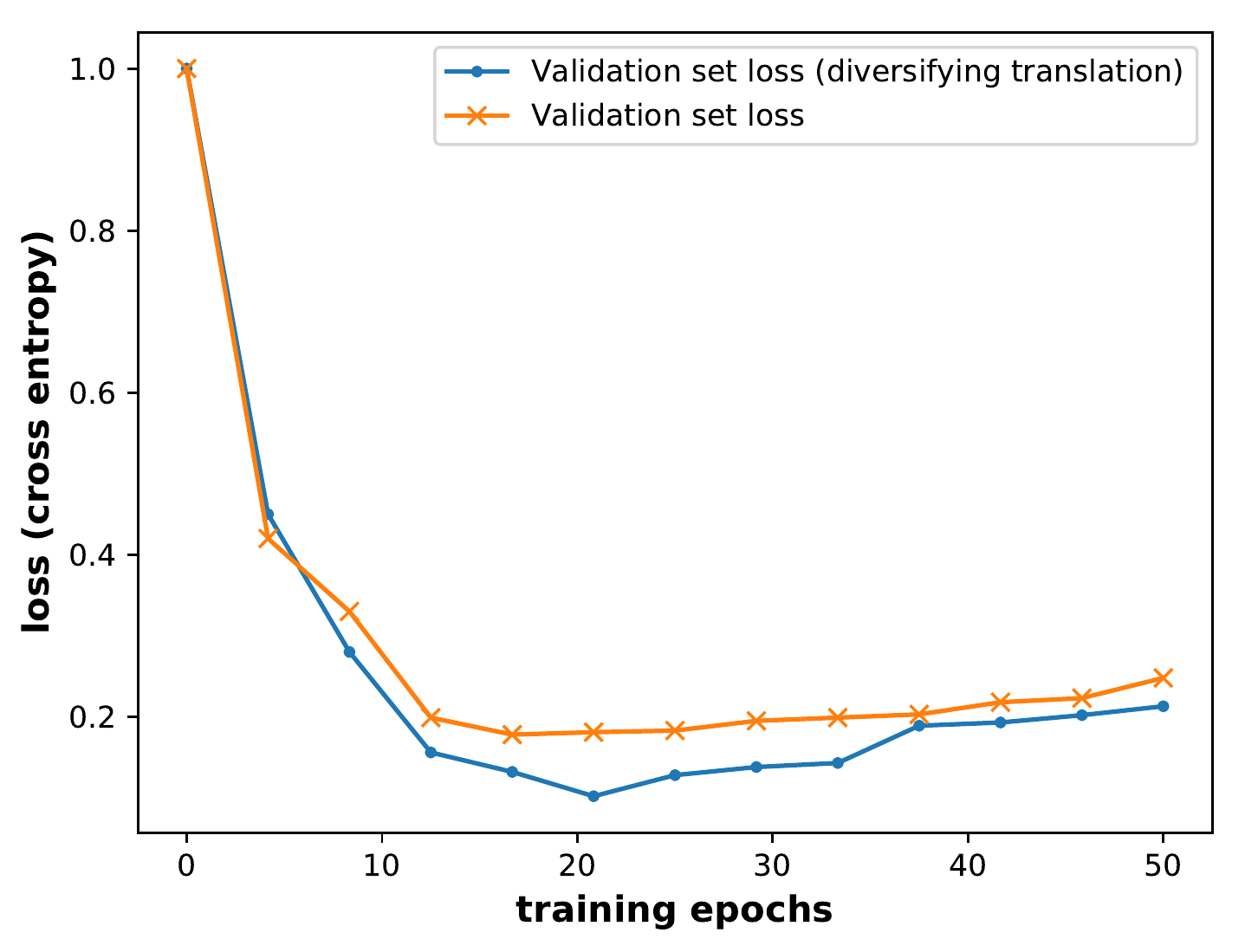}}
	\subfigure[Pre-trained word vectors]{
		\label{loss_impove}
		\includegraphics[width=0.7\columnwidth]{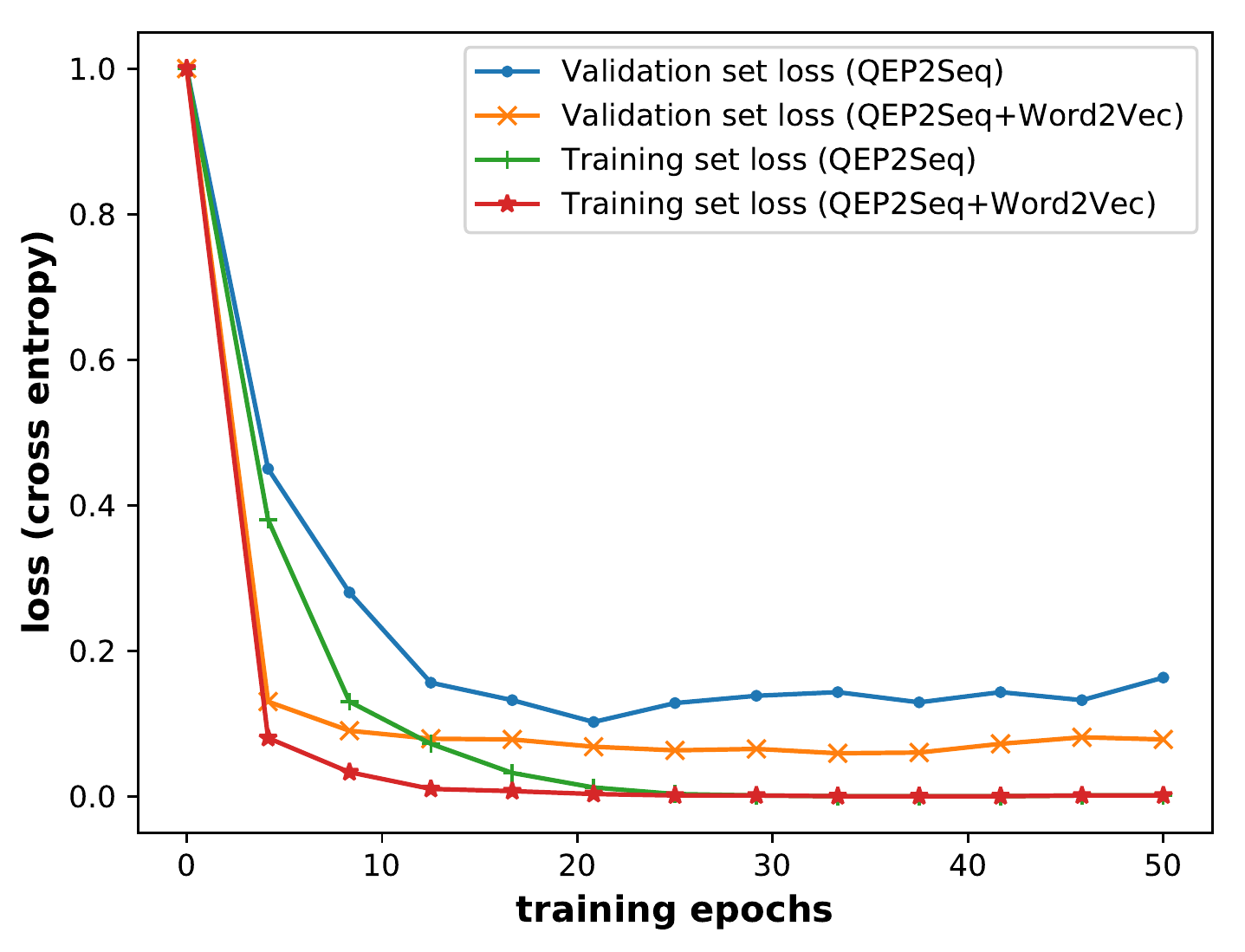}}
	\vspace{-1ex}\caption{Paraphrasing and pre-trained word vectors.}\label{fig8}
	\vspace{0ex}\end{figure}

\vspace{0ex}
\subsection{Experimental Setup}\label{sec:exptSetup}
\textbf{Datasets.} By default, we use PostgreSQL v10.12.2 as the underlying \textsc{rdbms}. Two \textsc{sme}s used \textsc{pool} to generate the descriptions of all physical operators to create the \textsc{poem} store. We use the \textsc{tpc-h} benchmark~\cite{tpch}, \textsc{sdss}~\cite{sdss}, and \textsc{imdb}~\cite{imdb} datasets as representatives of application domains\eat{(Table~\ref{tb:dataset})}. In particular, a recent benchmarking study~\cite{KS+20} reported that existing \textsc{nl2sql} techniques perform poorly on  \textsc{tpc-h} dataset containing complex and diverse \textsc{sql} queries.

We train our QEP2Seq model in \textsc{neural-lantern} using the workloads in \textsc{tpc-h} and \textsc{sdss}. 
For instance, given the 22 queries in \textsc{tpc-h}, we obtain their corresponding \textsc{qep}s. The \textsc{qep}s are then decomposed into 544 \emph{act}s. For each of them, we generate a series of natural language descriptions following the procedure described in Section~\ref{specialenvir}, resulting in 1632 samples. On the other hand, we generate 608 samples from the 71 predefined workload (\url{http://skyserver.sdss.org/dr16/en/help/docs/realquery.aspx}) in \textsc{sdss}. The neural network is implemented using Keras 2.2.4 and TensorFlow 1.13.2. We use \emph{Word2Vec}~\cite{WV},  \emph{GloVe}~\cite{glove}, \emph{ELMo}~\cite{elmo}, and \emph{BERT}~\cite{bert} as pre-trained word vectors in the \emph{Decoder}.

Note that \textsc{sdss} is tailored for SQL Server. Hence,  we implement \textsc{rule-lantern} (\textsc{neural-lantern} relies on QEP2Seq and is orthogonal to the underlying \textsc{rdbms}) on SQL Server (v15.0) as follows. First, \textsc{qep}s of SQL Server are in \textsc{xml} format. Hence, we implement a parser to transform a \textsc{qep} to the corresponding operator tree. Second, all physical operators of SQL Server are created using \textsc{pool} and stored in the \textsc{poem} store.\eat{ In summary, the differences between both implementations can be categorized into the followings: the same operator with different names (\eg \textsf{`Seq Scan'} in PostgreSQL and \textsf{`Table Scan'} in SQL Server); the operator uniquely in SQL Server (\eg \textsf{`Compute Scalar'}) and the one uniquely in PostgreSQL (\eg \textsf{`Materialize'}).} In summary, we can extend \textsc{lantern} to any \textsc{rdbms} easily by writing a parser to create operator trees and updating the \textsc{poem} store with \textsc{rdbms}-specific physical operators.

Finally, from all the samples, 80\% of them are randomly selected to train the model, while the remaining 20\% are selected as the validation set. Note that the performance of QEP2Seq is affected by the average number of training samples for each operator. In our experiments, there are on average 100 samples for each operator. 

\begin{table}[t]
	\begin{center} \scriptsize
		\begin{tabular}{|l|l|l|}
			\hline
			\textbf{Method} & \textbf{Self-BLEU} & \textbf{\#Samples per group} \\
			\hline
			\textit{Without paraphrasing}	& 1.0 & 1 \\
			\hline
			\textit{paraphrasing with}~\cite{synonymous3} & 0.309 & 2 \\
			\hline
				\textit{paraphrasing with}~\cite{synonymous2}& 0.603& 2 \\
			\hline
				\textit{paraphrasing with}~\cite{synonymous1}& 0.502& 2  \\
			\hline
			\textit{paraphrasing with}~\cite{synonymous1,synonymous2,synonymous3} & 0.482& 4\\
			\hline
		\end{tabular}
		\vspace{0ex} \caption{Diversity among the training samples.}\label{tb:self-bleu}
	\end{center}
	\vspace{-2ex}\end{table}

The trained model is applied to \textsc{imdb} for testing to demonstrate the portability of  \textsc{lantern} across different domains. Specifically, we generate 1000 \textsc{sql} queries using the approach in~\cite{DBLP:conf/cidr/KipfKRLBK19}. The corresponding \textsc{qep}s for these queries are then decomposed into 5232 acts, each of which is viewed as a test sample.


\vspace{1ex} \noindent \textbf{Algorithms.} We compare \textsc{lantern} with \textsc{neuron}~\cite{neuron}, a rule-based approach for generating natural language descriptions of \textsc{qep}s.  We also compare it with the textual and visual tree-based descriptions of \textsc{qep}s in PostgreSQL/SQL Server.

\begin{figure}[t]
	\centering
	\subfigure[pre-trained vs. self-trained]{
		\label{test_all}
		\includegraphics[width=0.7\columnwidth]{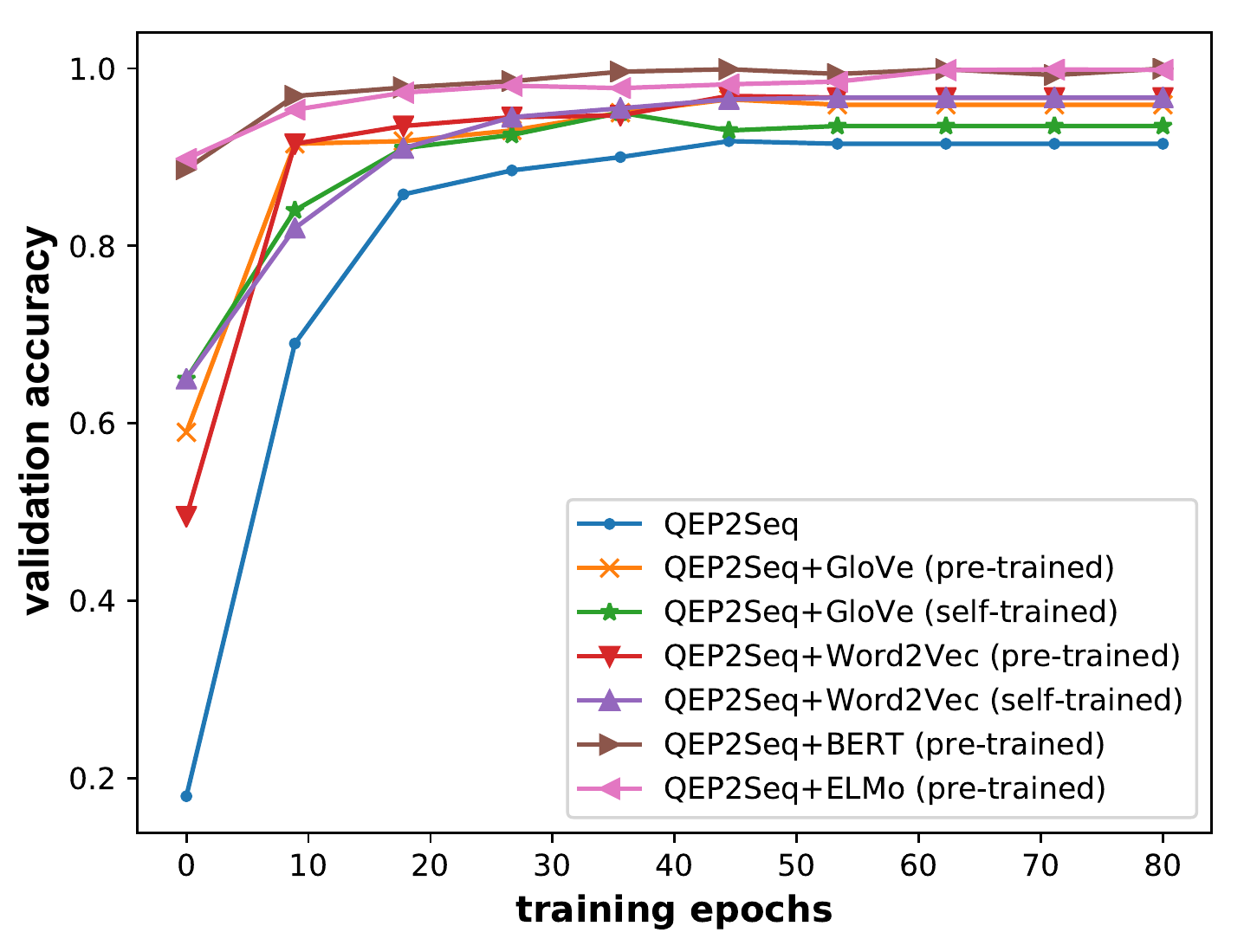}}
	\subfigure[w or w/o learned parameters]{
	\label{test_train_and_no_train}
	\includegraphics[width=0.7\columnwidth]{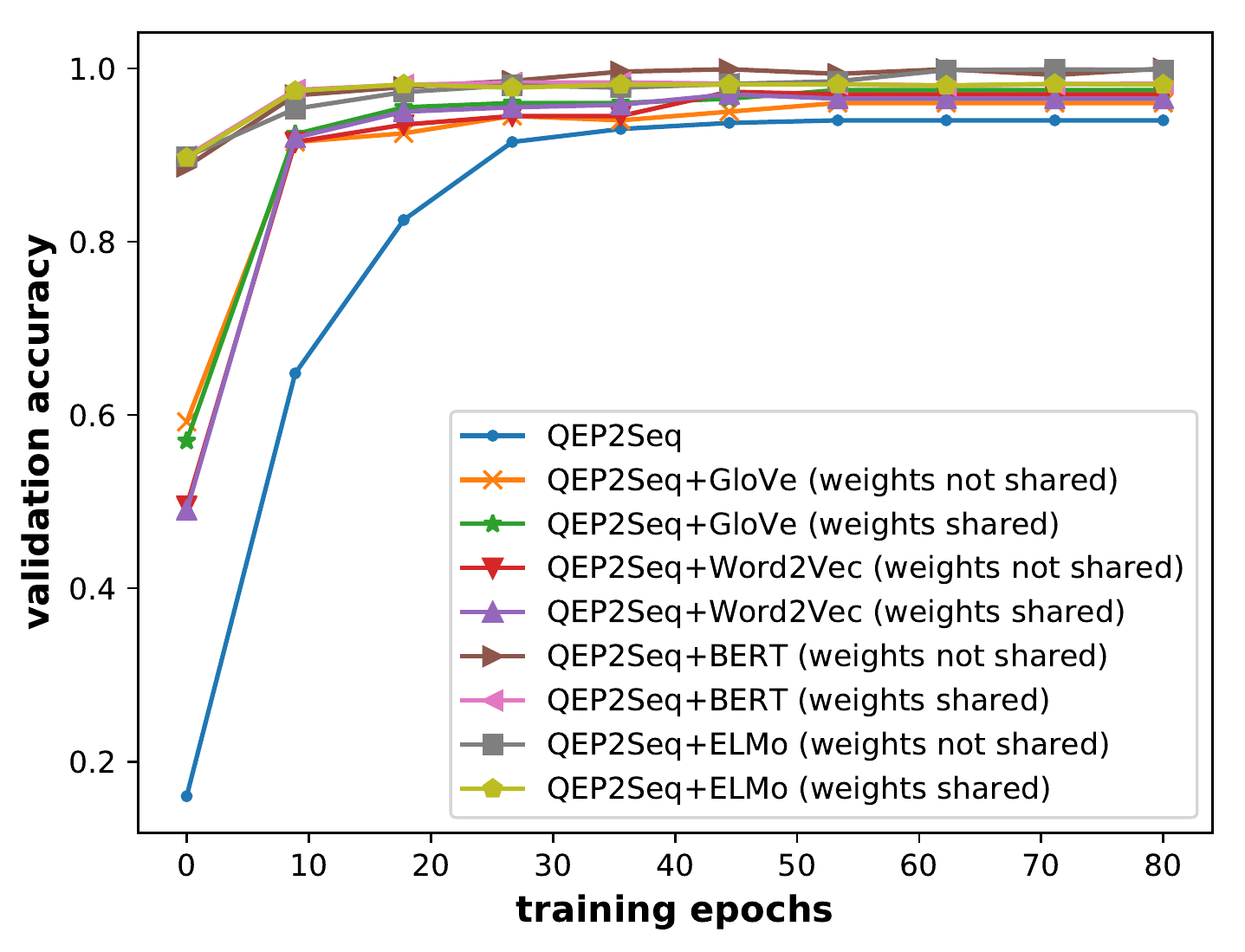}}
	\vspace{-1ex}\caption{Impact of pre-trained vectors.}\label{fig9}
	\vspace{0ex}\end{figure}

\begin{figure*}[t]
   \centering
          \subfigure[Length of input vs. output.]{
          \label{fg:length}
          \includegraphics[width=0.8\linewidth]{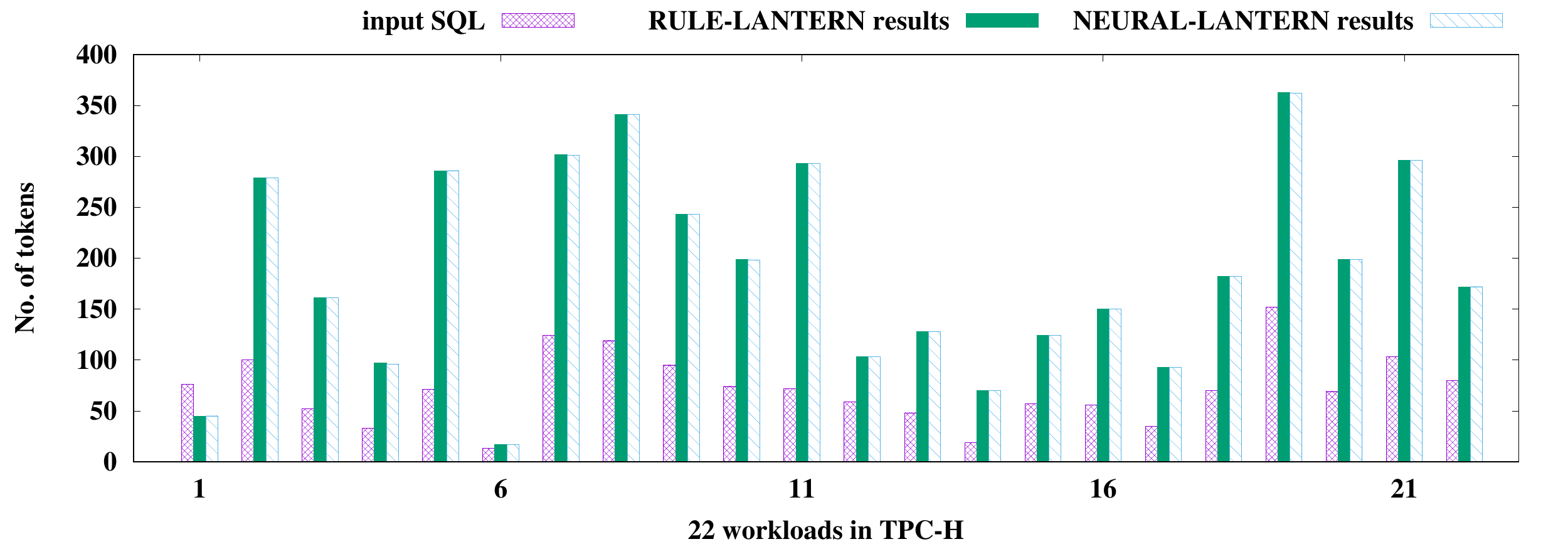}}
   \subfigure[Question Q1]{
          \label{q1cn}
          \includegraphics[width=0.3\linewidth]{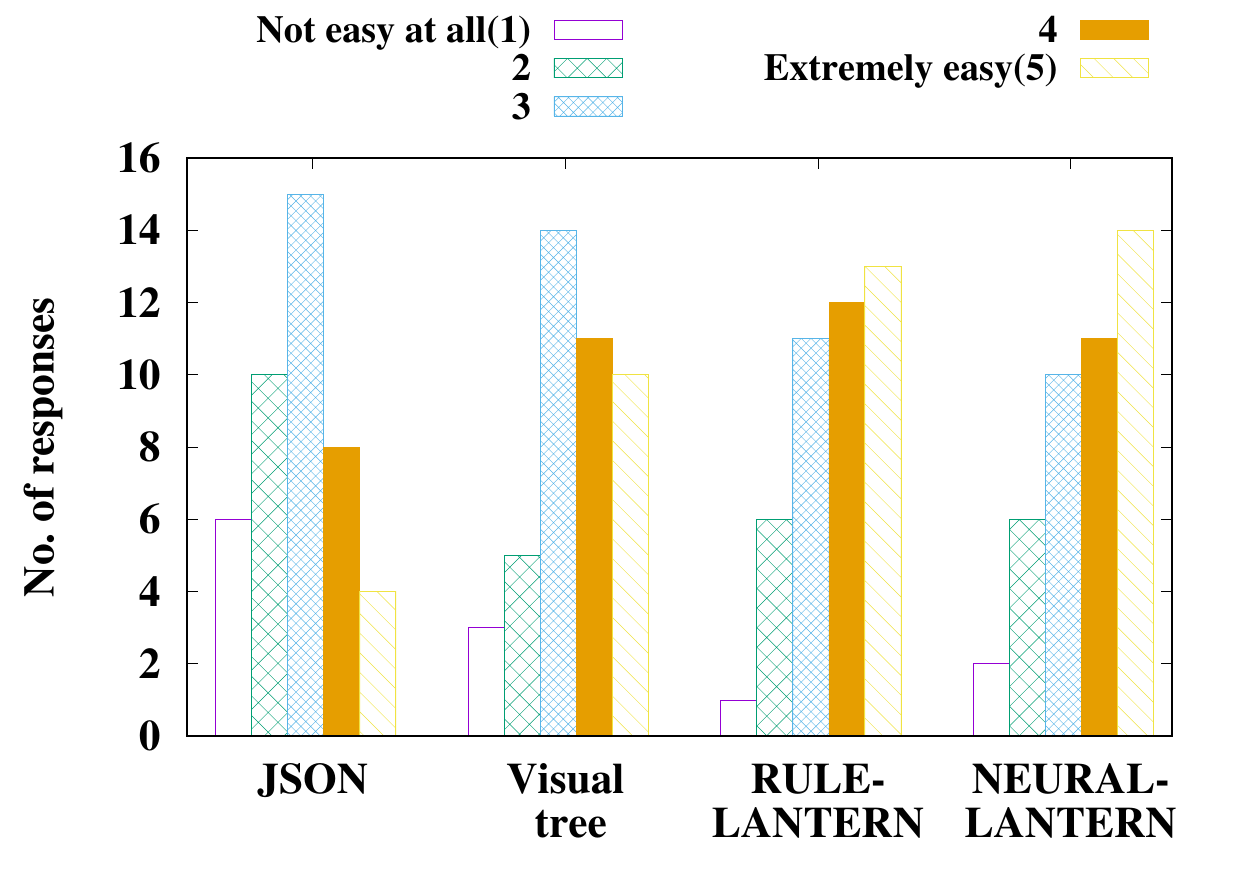}}
   \subfigure[Question Q2]{
          \label{q2sg}
          \hspace{-3ex} \includegraphics[width=0.3\linewidth]{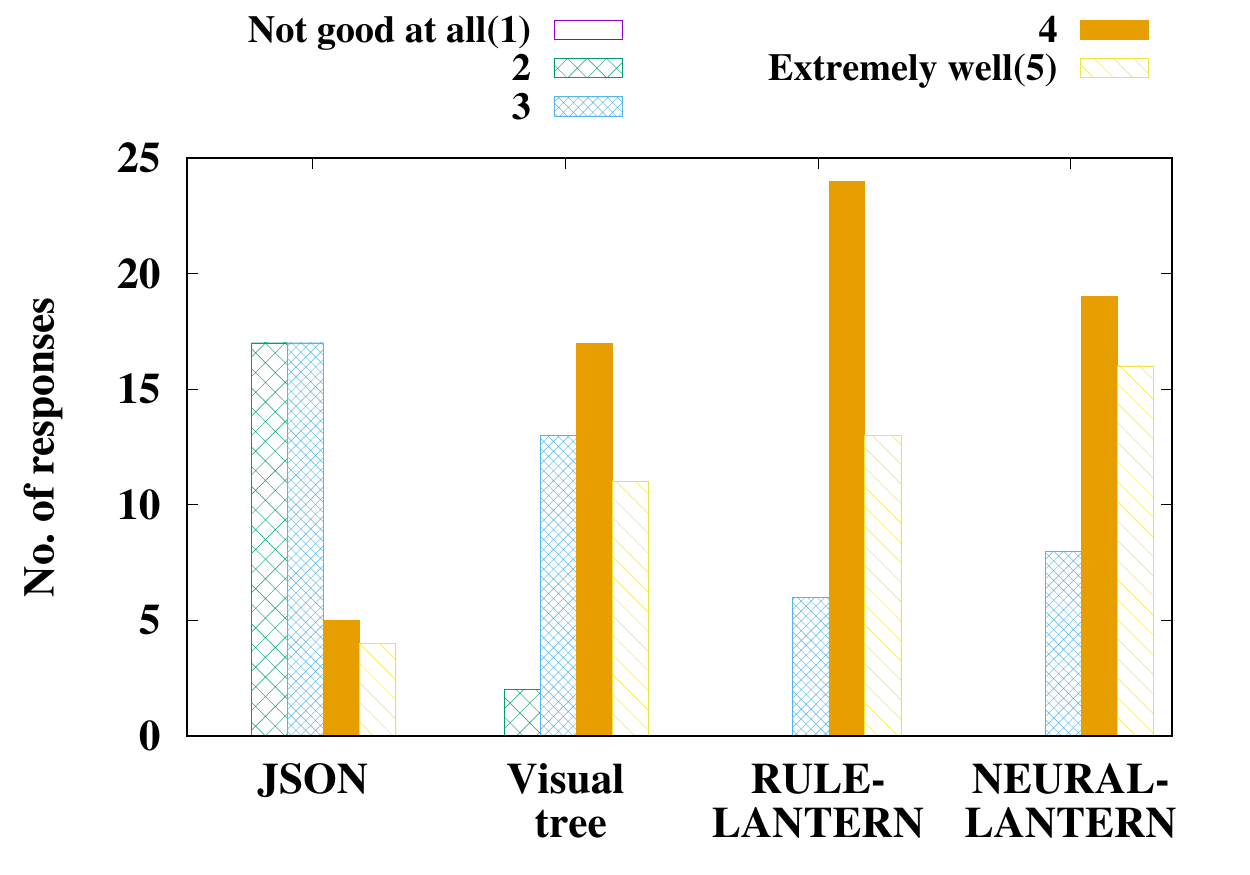}}
   \subfigure[Question Q3]{
          \label{q3cn}
          \hspace{-3ex} \includegraphics[width=0.3\linewidth]{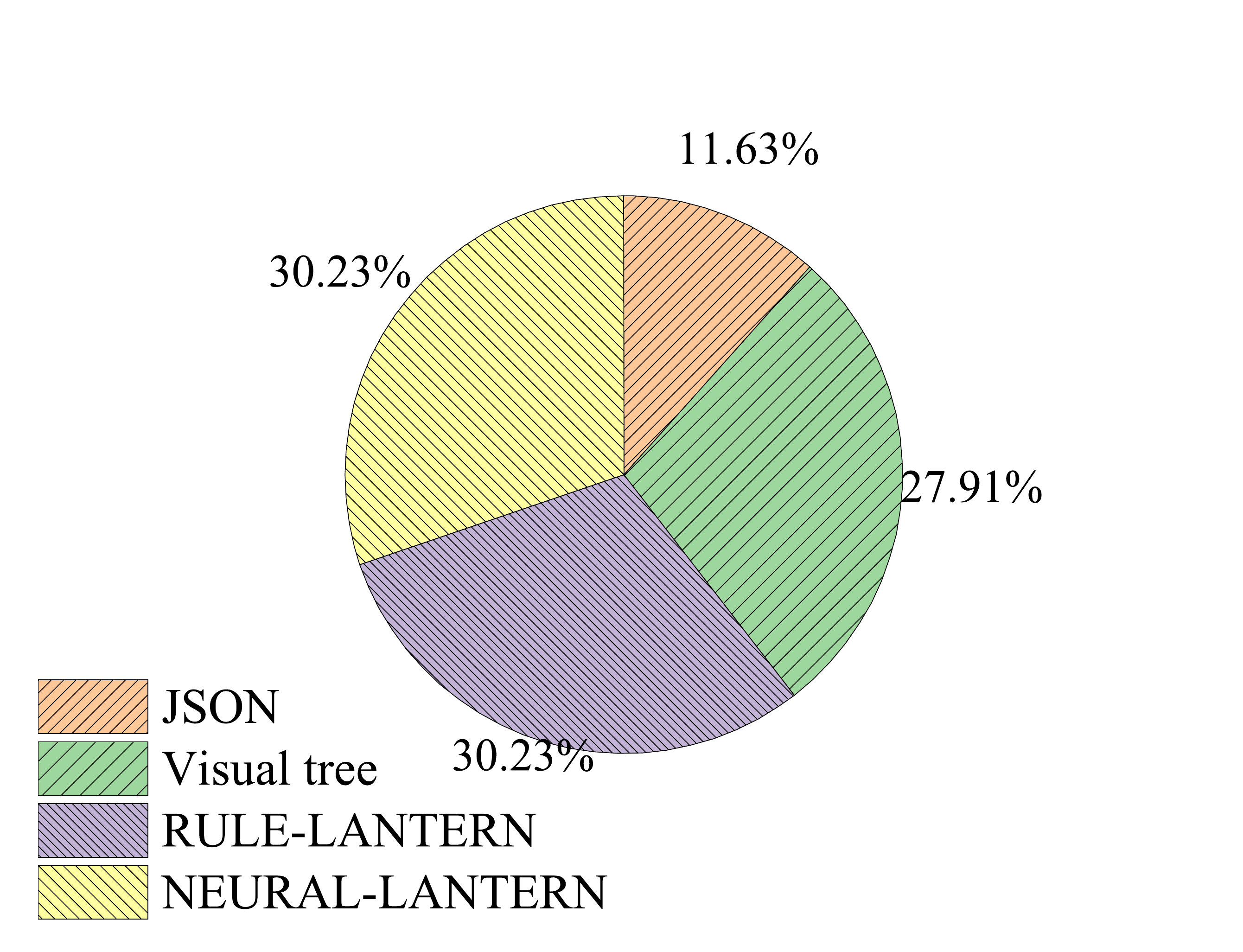}}
   \vspace{-2ex}\caption{(a) Length of output; (b)-(d) Responses to Question 1-3.}\label{fig:userstudyq1}
   \vspace{0ex}\end{figure*}

\vspace{0ex}
\subsection{Experimental Results}\label{ssec:expresult}

\textbf{Exp 1: Effect of diversifying text.} First, we report the benefits brought by paraphrasing in \textsc{neural-lantern}.\eat{ Accordingly, the original training set is enlarged (\ie from 544 samples to 1632 in \textsc{tpc-h}).}  Table~\ref{tb:self-bleu} reports the \textit{diversity} of \textsc{nl} descriptions measured using \textit{Self-BLEU}~\cite{SNC+19} (normalized to $0\sim 1$, a lower value indicates a higher diversity), which is widely adopted in machine translation tasks to measure diversity of the generated text in a language.  Given the 1152 samples (544 from \textsc{tpc-h}, 608 from \textsc{sdss}) generated by \textsc{rule-lantern}, we apply the paraphrasing tools over each sample. As a result, each original sample (from \textsc{rule-lantern}) as well as its variations (generated by paraphrasing) form a \textit{group}. We compute the diversity of each group and report the average over all 1152 groups.
Notably, \textit{\#Samples per group} column refers to the number of samples in each group. Recall that we eliminate invalid or duplicate paraphrasing results. Hence, a few groups may have fewer samples than the theoretical values listed in the column. Clearly, paraphrasing is beneficial w.r.t diversity of \textsc{nl} descriptions. Next, we use the diversified samples (\ie paraphrasing with~\cite{synonymous1,synonymous2,synonymous3}) for training and evaluate the validation loss (\ie cross entropy loss in Eq~\ref{logloss}). Figure~\ref{data_augmentation} plots the results. Observe that paraphrasing reduces the loss significantly.

\vspace{1ex}\noindent\textbf{Exp 2: Length of output.} Although paraphrasing enables \textsc{neural-lantern} to generate descriptions with high diversity and low validation loss, does it make the descriptions verbose? We now report  lengths of the \textsc{nl} descriptions generated by \textsc{rule-lantern} and \textsc{neural-lantern} to answer this. Figure~\ref{fg:length} plots the lengths of the original \textsc{sql} statements in \textsc{tpc-h} as well as outputs of these two techniques. We observe that the length of a natural language description is, in fact, affected by the complexity and the number of relations in a \textsc{sql} statement, but not by the length of the statement. Importantly,  \textsc{neural-lantern} injects variability without adversely impacting the length significantly compared to \textsc{rule-lantern}.

\vspace{1ex}\noindent\textbf{Exp 3: Effect of pre-trained word embeddings.} Next, we compare the changes to the loss function by employing \textit{Word2Vec} or otherwise. Figure~\ref{loss_impove} depicts the results.  Observe that the adoption of pre-trained word vector can speed up training and significantly reduce the validation loss. We also notice that during training, the training set loss first decreases and then slowly increases (over 35 epochs). Hence we apply an early stopping strategy to prevent overfitting.  Specifically, we terminate training when the training set loss fluctuation range is less than a threshold (\eg 0.001).

\vspace{1ex}\noindent \textbf{Exp 4: Varying the pre-trained word vectors.} We conduct a set of experiments to test the performance of \textsc{neural-lantern} by varying the pre-trained word vectors. In particular, we compare the following five approaches: \textit{QEP2Seq} (with randomly initialized word embeddings), \textit{QEP2Seq}+\textit{GloVe}, \textit{QEP2Seq}+\textit{Word2Vec}, \textit{QEP2Seq}+\textit{BERT}, and \textit{QEP2Seq}+\textit{ELMo}. Figure~\ref{test_all} depicts the results in terms of \textit{sparse\_categorical\_accuracy}~\cite{keras} averaged over all output sequences. For each output sequence with $m$ tokens, it can be calculated as $Acc = \frac{1}{m}\sum_{t=1}^m \mathbbm{1} (y_t=o)$. Observe that the training process is faster and the accuracy on the development set is higher when pre-trained vectors are adopted. As expected, the performance for the contextual embeddings (\textit{ELMo}, \textit{BERT}) are the best. In addition, using existing pre-trained word vectors, which are trained on large corpus such as Wikipedia, show superior results to our self-trained word vectors (referred to as \textit{self-trained} in the figure), which are pre-trained on our \textsc{rule-lantern} output. This is expected as the dataset size is limited for the latter.

We also compare the impact of sharing and not sharing the weights between the \textit{Encoder} and \textit{Decoder}. The results are shown in Figure~\ref{test_train_and_no_train}. Observe that the performances are comparable for models with pretrained embeddings. 

Lastly, in line with existing Seq2Seq models, we adopt a measure that is widely used in machine translation task, \ie \emph{BLEU} \cite{papineni2002bleu}. For each specific approach of \textsc{neural-lantern}, we compute the \emph{BLEU} score of its output with respect to the ground-truth and report the average over all samples. The results are presented in Table~\ref{tb:bleu}. Clearly, \textit{QEP2Seq}+\textit{BERT} demonstrates the most similar results with respect to the ground-truth. \eat{For reference, we list down a subset of the SQL statements as well as their \textsc{qep}s in Appendix~\ref{app:output}, where the generated NL translation from our \textsc{neural-lantern} are also included.}

\begin{table}[t]
	\begin{center} \scriptsize
		\begin{tabular}{|l|l|}
			\hline
			\textbf{Method} & \textbf{BLEU score (test set)} \\
			\hline
			\textit{QEP2Seq}	& 51.46 \\
			\hline
			\textit{QEP2Seq}+\emph{GloVe} (pre-trained) & 68.15 \\
			\hline
			\textit{QEP2Seq}+\emph{GloVe} (self-trained)& 57.01 \\
			\hline
			\textit{QEP2Seq}+\emph{Word2Vec} (pre-trained)& 64.01  \\
			\hline
			\textit{QEP2Seq}+\emph{Word2Vec} (self-trained) & 54.85\\
			\hline
			\textbf{\textit{QEP2Seq}+\emph{BERT} (pre-trained)} & \textbf{73.73}\\
			\hline
			\textit{QEP2Seq}+\emph{ELMo} (pre-trained) & 71.67\\
			\hline
		\end{tabular}
		\vspace{0ex} \caption{The performance of \textit{QEP2Seq} (with beam size 4)}\label{tb:bleu}
	\end{center}
\vspace{0ex}\end{table}

\begin{table}[t]
	
	\begin{center} \scriptsize
		\begin{tabular}{|p{.6cm}|p{1.2cm}|p{1.3cm}|p{1cm}|p{1.2cm}|p{1.2cm}|}
			\hline
			\textbf{Steps} & \textbf{Training} (over \textsc{tpc-h} and \textsc{sdss} samples) & \textbf{Training} for each epoch & SQL generation (1000 queries in \textsc{imdb}) & \textsc{neural-lantern} avg. response time & \textsc{rule-lantern} avg. response time\\
			\hline
			\textbf{Time} & 825.60  & 16.51 [18.22]  & 0.77 & 0.216 & 0.015\\
			\hline
		\end{tabular}
		\vspace{-1ex}\caption{Efficiency (in sec).}\label{tb:time}
	\end{center}
\vspace{0ex}\end{table}

\vspace{1ex}\noindent \textbf{Exp 5: Errors in \textsc{neural-lantern}.} \eat{The aforementioned experiments report the effectiveness (\resp deviation) for the translated description compare to the \textsc{rule-lantern} output.} Observe that neither \textit{accuracy} nor \textit{BLEU} can justify the correctness (\ie whether the translation make sense for human understanding) of the output of \textsc{neural-lantern}. Hence, we employ two \textsc{sme}s to manually check the correctness of the \textsc{nl} descriptions.  We uniformly sample 100 test samples randomly, and test whether the translated descriptions are correct. We find that 83 are correctly translated; another 13 has one wrong token; the remaining four contains 6-9 wrong tokens. In the next section, we shall investigate the impact of these descriptions on facilitating understanding of \textsc{qep}s among learners.

\vspace{1ex}\noindent\textbf{Exp 6: Efficiency.} Table~\ref{tb:time} reports the time cost of different components. Firstly, the average training time for each epoch is 16.51 (resp. 18.22) sec for  $QEP2Seq+GloVe$ (resp. $ QEP2Seq+ELMo$), which has the least (resp. largest) number of dimensions. Secondly, the average time taken to  generate a \textsc{nl} description is less than a second.
\vspace{0ex}
\subsection{User Study}
We conducted a  user study among \textsc{cs} undergraduate students who are taking the database course in an institution.  43 unpaid volunteers participating in the study. We utilized the \textsc{gui} of \textsc{neuron}~\cite{neuron} for presenting the input queries and output translations of \textsc{lantern}. Rest of the features (\eg question answering module) of \textsc{neuron} are orthogonal to this work and hence were disabled.  We presented a 10-min scripted tutorial of the \textsc{gui} describing how to use it. We then allowed the subjects to play with the tool for 15 min.\eat{ During this time we answered any questions they had about the \textsc{gui}.}

\vspace{1ex}\noindent\textbf{US 1: Survey.} Each of the subjects was given the \textsc{qep}s corresponding to the queries in \textsc{tpc-h}/\textsc{sdss} and their natural language descriptions generated by \textsc{rule-lantern} and \textsc{neural-lantern}. The subjects were not informed on which description was generated by which technique. They were also given the corresponding \textsc{json}/\textsc{xml} and visual tree formats of \textsc{qep}s generated by PostgreSQL/SQL Server. The queries as well as outputs of different approaches were given in random order.   They were allowed to take their own time to peruse the plans. After the completion of the task, the subjects were asked to fill up a survey form, which consists of a series of questions to understand the impact of various modes of \textsc{qep} on their understanding of how \textsc{sql} queries are executed. They were instructed that the outcomes of the survey have no bearing on their course grades. We now elaborate on the key results from the survey  on \textsc{tpc-h} (results on \textsc{sdss} are qualitatively similar).

\eat{\begin{figure}[t]
	\centerline{\includegraphics[width=\columnwidth, height=2.6cm]{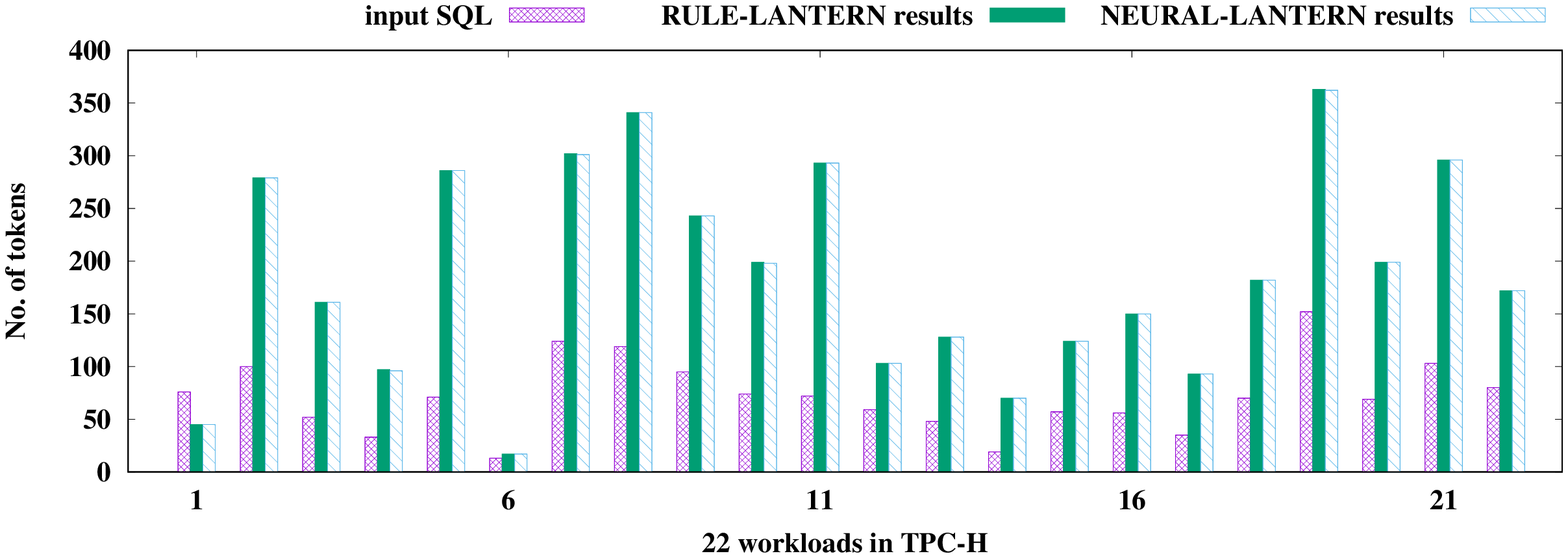}}
	\vspace{-3ex}\caption{Length of input vs. output.}
	\label{fg:length}
\vspace{-3ex}\end{figure}

\begin{figure}[t]
	\centering
	\subfigure[Question Q1]{
		\label{q1cn}
		\includegraphics[width=0.51\columnwidth, height=2.7cm]{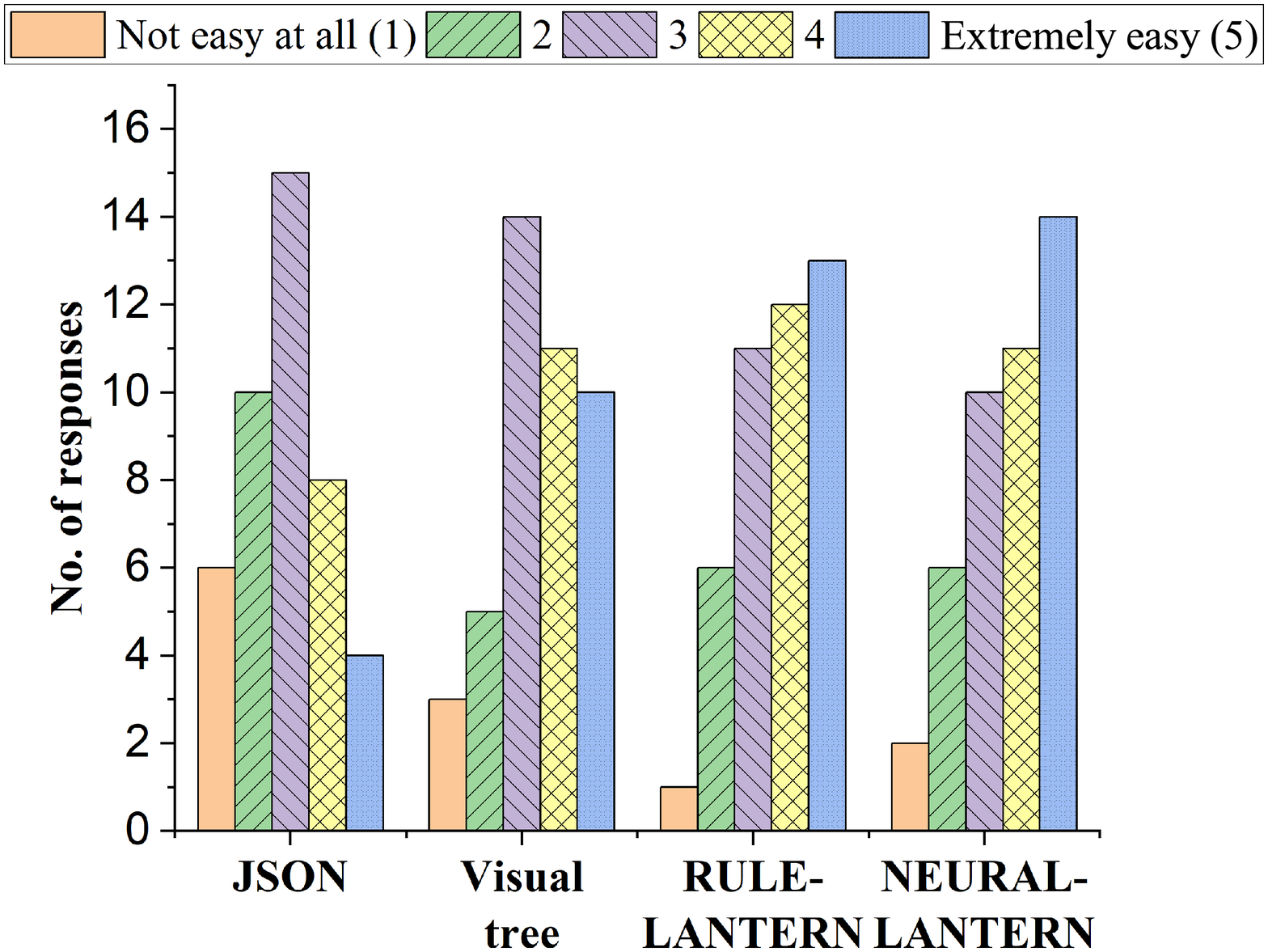}}
	\subfigure[Question Q2]{
		\label{q2sg}
		\hspace{-3ex} \includegraphics[width=0.49\columnwidth, height=2.7cm]{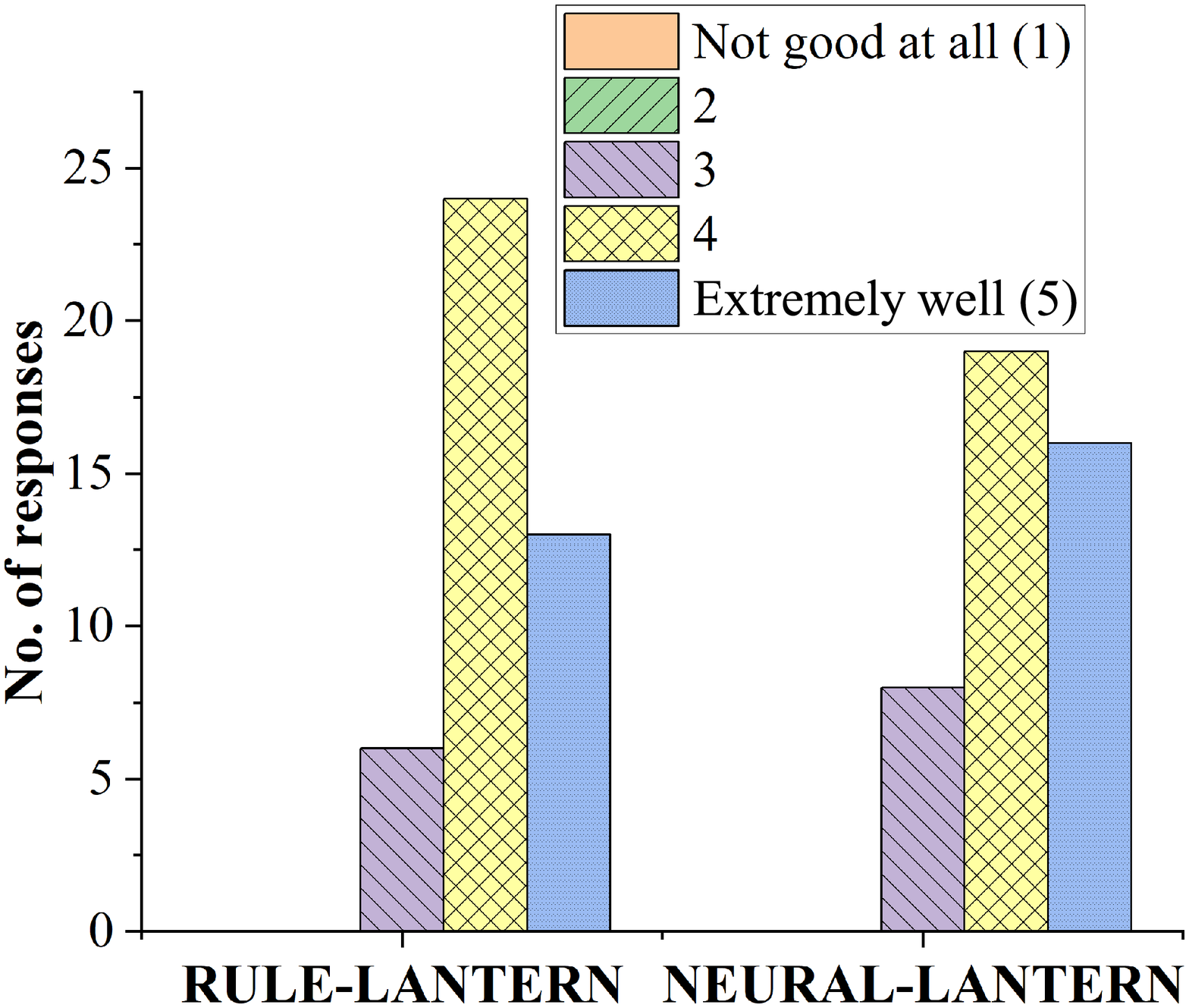}}
	\vspace{-3ex}\caption{Responses to Question 1 and 2.}\label{fig:userstudyq1}
	\vspace{-3ex}\end{figure}}

\textit{Q1: How easy it is to understand the query plan presented using each approach?} Figure~\ref{q1cn} shows the statistics with respect to the number of responses for this question. Each subject gave a rating in the Likert scale of 1-5. Observe that the \textsc{lantern} approach is the easiest format to understand. In particular, for both \textsc{rule-lantern} and \textsc{neural-lantern}, 58.1\% of ratings are above 3 for both solutions. In comparison, there are 27.9\% and 48.8\% ratings above 3 for \textsc{json} and visual tree, respectively. Note that majority of the volunteers (41 out of 43) did not raise any issue with the length of the \textsc{nl} descriptions generated by \textsc{lantern}.

\textit{Q2: How well does \textsc{lantern} describe the query plans?} Figure~\ref{q2sg} reports  the results.  86\% (\resp 81.4\%) of the respondents agree that the \textsc{rule-lantern} (\resp \textsc{neural-lantern}) does a good job in describing the query plans to facilitate understanding of query execution steps. Slightly higher agreement for the former is expected as hand-written rule-based technique is expected to be more accurate than the neural-based approach. We also observe that there is no significant impact of different pre-training models employed in \textsc{neural-lantern} (Figure~\ref{pretrainus}). This is not surprising as given the constrained nature of the problem (\ie both input and output), large pretrained models like BERT has little scope to improve qualitatively.

\eat{We also ask the volunteers to compare between \textsc{neural-lantern} of different pre-trained models. In particular, the output of different models are often the same for simple workloads we generated following~\cite{DBLP:conf/cidr/KipfKRLBK19}. Therefore, in this study the volunteers are asked to use only \textsc{tpc-h} workloads, the interpretations of whose \textsc{qep} differentiate enough under our manual check. As shown in Figure~\ref{pretrainus}, there is no significant difference among the pre-training models employed in \textsc{neural-lantern}. This is not surprising as given the constrained nature of the problem (\ie both input and output), large pretrained models like BERT has little scope to improve qualitatively.}

We also study whether the diverse descriptions generated by \textsc{lantern} are confusing the volunteers. To this end, we generate 20 pairs of \textsc{nl} descriptions. 10 of these are positive examples where each pair is associated with the \textit{same} \textsc{sql} query. That is, in each pair, the two descriptions  are generated by \textsc{rule-lantern} and \textsc{neural-lantern} for the same \textsc{qep}. The remaining 10 pairs are negative examples where the two \textsc{nl} descriptions in each pair are from two \textit{different}  \textsc{qep}s. The 20 pairs are then given to the volunteers in random order and they were tasked to identify the positive example pairs. All volunteers correctly identified all the positive pairs.

\textit{Q3: Which query plan format do you prefer the most?} Figure~\ref{q3cn} reports the results. Both solutions of \textsc{lantern} are preferred the most. In particular, \textsc{rule-lantern} and \textsc{neural-lantern} receive similar preferences. On the other hand, very few, \ie 11.63\%, participants chose the textual format as the most preferred choice.

\vspace{1ex}\noindent\textbf{US 2: Impact of paraphrasing.} We now conduct a user study to test the impact of incorporating paraphrasing in \textsc{neural-lantern}. The participants answer $Q2$ again but now they study the outputs of \textsc{neural-lantern} with (w) and without (w/o) usage of paraphrasing. The results are shown in Figure~\ref{paraus}. The user experience of \textsc{neural-lantern} w/o paraphrasing is worse than with paraphrasing. In fact, when we eliminate the samples generated by paraphrasing, the results of \textsc{neural-lantern} contain many error tokens (\eg missing filtering conditions) due to limited number of training samples and overfitting problem.

\begin{table}[t]
	\begin{center} \scriptsize
		\begin{tabular}{|c|p{0.8cm}|p{0.8cm}|p{0.8cm}|p{0.8cm}|p{0.8cm}|}\hline
		\multirow{3}*{\textbf{Method}}	 & \multicolumn{5}{|c|}{\textbf{Boredom index}}\\
			& \multicolumn{5}{|c|}{(not boring $\rightarrow$ extremely boring)} \\
			\cline{2-6}
			&\textbf{1} & \textbf{2}&\textbf{3}&\textbf{4}&\textbf{5} \\
			\hline
			\textsc{rule-lantern}	& 2&7&19&10&5 \\
			\hline
			\textsc{neural-lantern} & 6&11&22&3&1 \\
			\hline\hline
			\textsc{neuron} & 2&8&16&11&6 \\
			\hline
			\textsc{lantern} & 6&12&21&2&2 \\
			\hline
		\end{tabular}
			\vspace{-1ex}\caption{Impact on boredom.}\label{tb:boredum}
	\end{center}
\vspace{-3ex}\end{table}

\vspace{1ex}\noindent\textbf{US 3: Impact of habituation and boredom.} The diversified translation of \textsc{neural-lantern} aims to mitigate the potential boredom suffered by subjects in using \textsc{rule-lantern}. To validate this issue, we presented a set of output generated by each approach in random order and asked the subjects to rate the degree of boredom (\ie \textit{ boredom index}) they felt perusing these plans to understand \textsc{qep}s using the Likert scale of 1-5 (1 refers to no boredom and 5 refers to the highest degree of boredom). Note that boredom literature relies heavily on subjective self-report measures~\cite{MF+11}. Table~\ref{tb:boredum} reports the results (first two rows). 15 (gave scores above 3) out of 43 volunteers felt that the output of \textsc{rule-lantern} makes them bored and prone to skipping text due to repeated exposure of the same descriptions over multiple workloads. In comparison, only 4 volunteers (\ie 9.3\%) felt results of \textsc{neural-lantern} are boring.

In the above settings, the subjects perused the outputs of \textsc{rule-lantern} and \textsc{neural-lantern} separately in random order. In this experiment, we randomly mix the results of the two. In particular, we adopt 50 \textsc{sql} statements generated by~\cite{DBLP:conf/cidr/KipfKRLBK19} over \textsc{imdb}, each of which contains \textsf{Hash Join} and \textsf{Aggregate} operators. We use \textsc{neural-lantern} to generate every $4+f()$ output, where $f()$ is a pseudorandom function with uniform probability to output $\{-1,0,1\}$. Others are generated using \textsc{rule-lantern}. As a result, the volunteers are given 14 \textsc{nl} descriptions from \textsc{neural-lantern}, mixed with 36 output generated by \textsc{rule-lantern}. They are unaware of which output is generated by which technique. They were asked to mark outputs that make them feel bored to peruse and those which arouse their interests without compromising on understanding the \textsc{qep}s. We observe that not all descriptions are marked w.r.t boredom. Particularly, 21  (resp. 14) descriptions generated by \textsc{rule-lantern} (resp. \textsc{neural-lantern}) are marked. Out of them 2 (resp. 8) descriptions of  \textsc{rule-lantern} (resp. \textsc{neural-lantern}) aroused interest. \eat{  82\% of outputs that aroused interest are generated by \textsc{neural-lantern}; while 95\% of those marked as boring are generated by \textsc{rule-lantern}.} In summary, our proposed neural approach indeed alleviates the impact of boredom on learners.

\begin{figure}[t]
	\centering
	\subfigure[Pre-trained models]{
	\label{pretrainus}
	\includegraphics[width=0.7\linewidth]{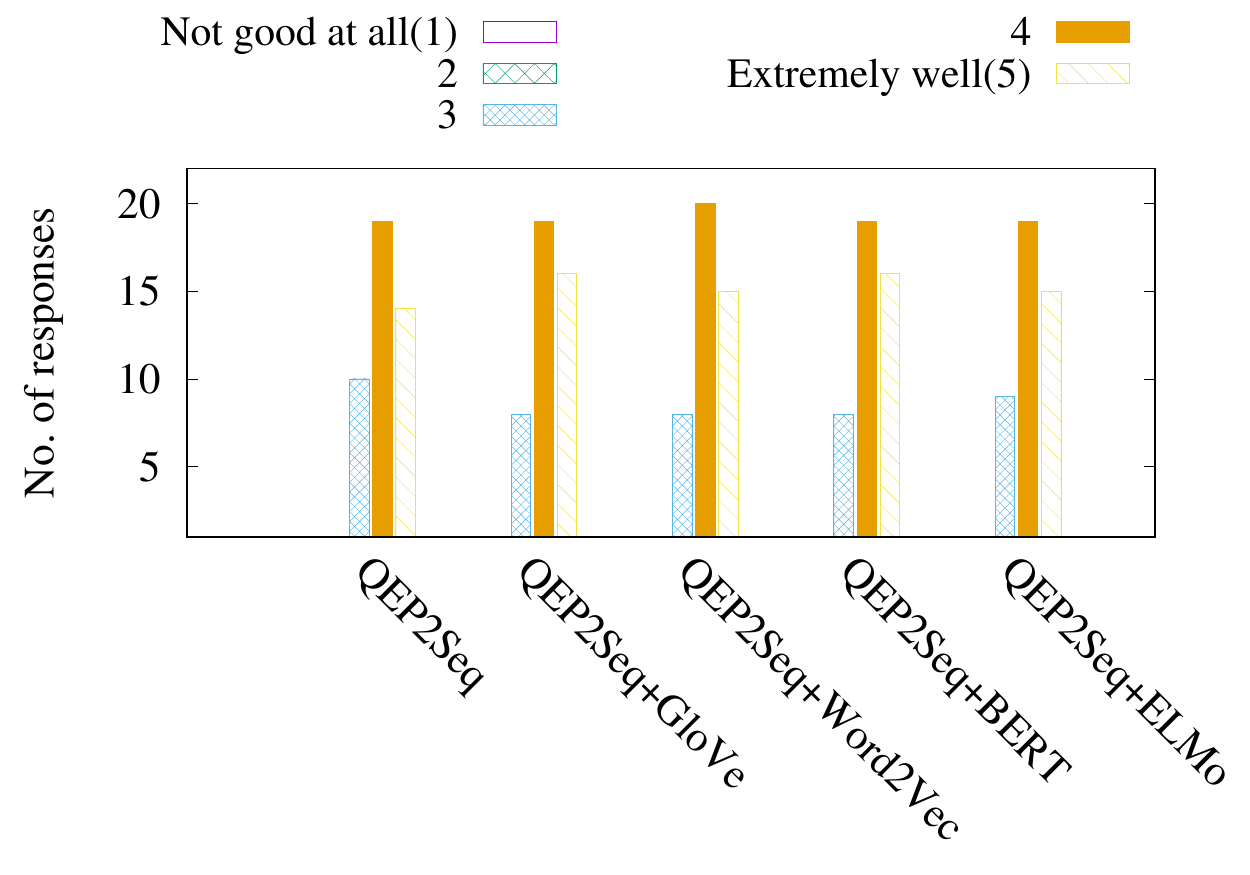}}
	\subfigure[Impact of paraphrasing.]{
		\label{paraus}
		\includegraphics[width=0.7\linewidth]{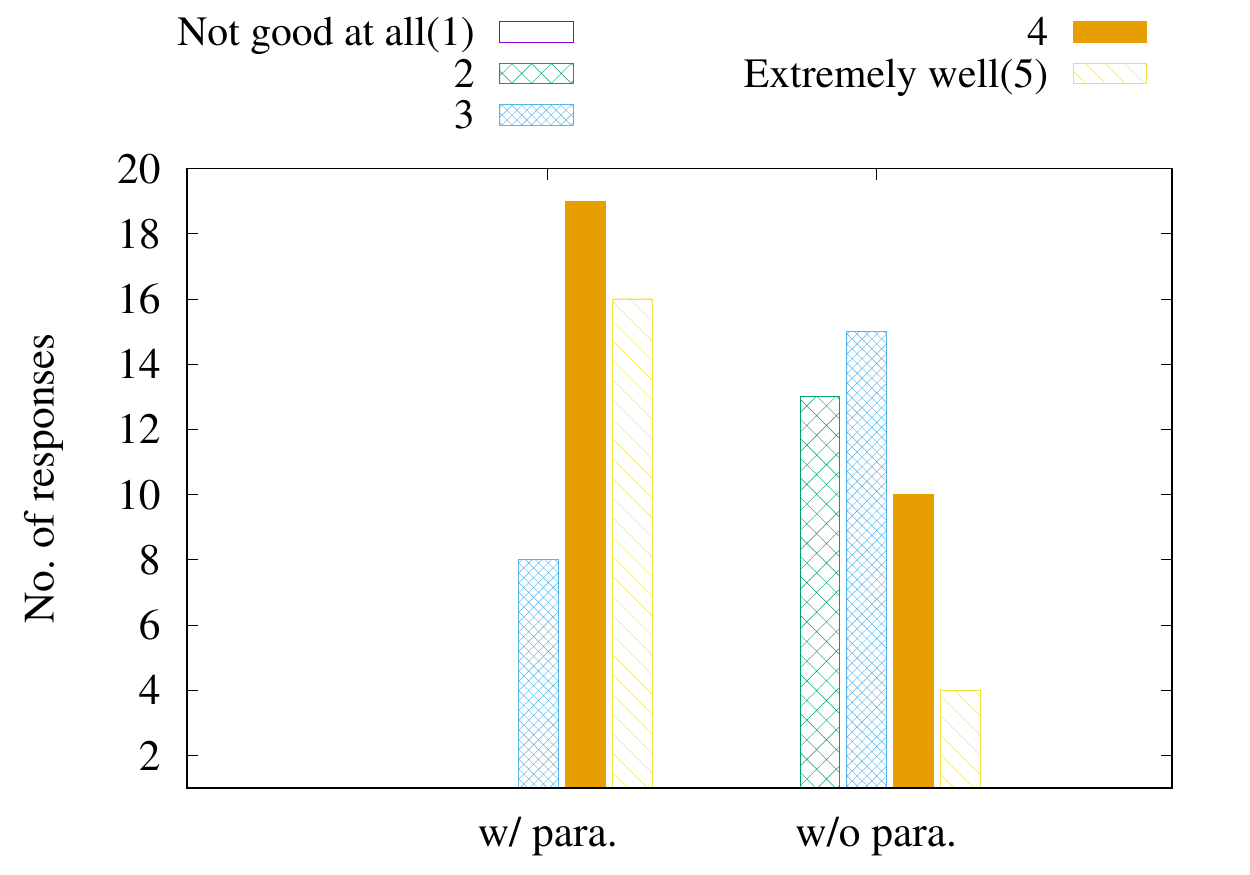}}
	\subfigure[\textsc{lantern} vs. \textsc{neuron}.]{
		\label{qsuppus}
		\includegraphics[width=0.7\linewidth]{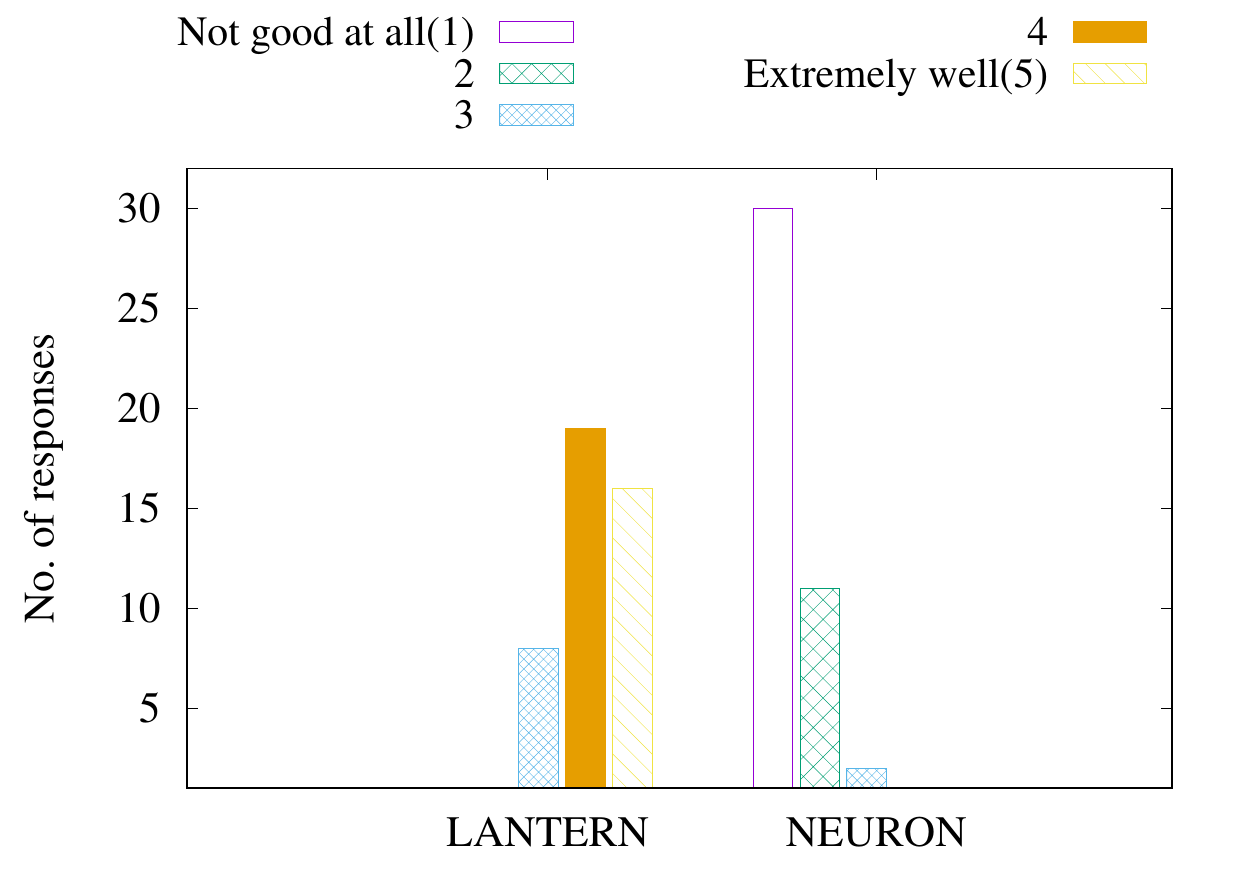}}
	\vspace{-1ex}\caption{User study (contd.)}\label{fig:userstudyq45}
	\vspace{0ex}\end{figure}

\eat{Besides, we also test whether the output \textsc{neural-lantern} differentiates from that of \textsc{rule-lantern} too much such that the user is confused. To answer the question, we randomly select 20 SQL queries, each of which is associated with a pair of NL descriptions by \textsc{rule-lantern} and \textsc{neural-lantern}, respectively. We show 20 pairs of the descriptions (but not the query statement), 10 of them are associated to the same query respectively and the other 10 are randomly coupled, to the volunteers and ask them to identify all those pairs corresponding to the same query. All the participants have made the correct choices. That is, although \textsc{neural-lantern} introduces diversity comparing with \textsc{rule-lantern}, it does not confuse the users at all.}

\begin{figure*}[t]
\centering
\includegraphics[width=0.8\linewidth]{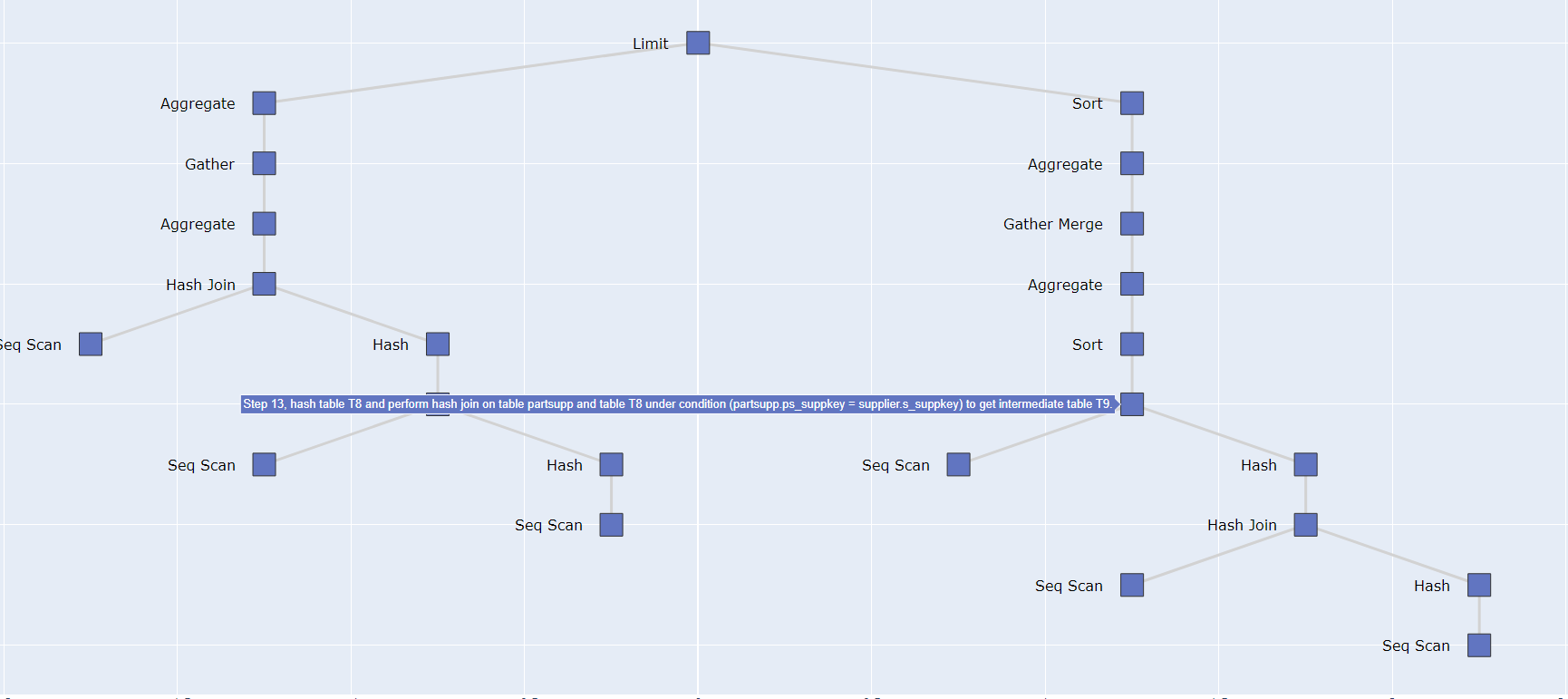}
\vspace{-2ex}\caption{An example of visual tree-based \textsc{nl} presentation format.}
\label{fig:vt-nl}
\vspace{0ex}\end{figure*}

\textbf{US 4: Impact of incorrect token on comprehension.} Recall that the \textsc{neural-lantern} may produce some wrong tokens  in the test samples. Hence, we ask the volunteers to evaluate whether the existence of wrong tokens affect their understanding of \textsc{qep}s or mislead their understandings for the corresponding operators. Our study revealed that only 2 out of 43  think that the incorrect tokens are problematic for their understanding (gave a rating below 3).

\vspace{1ex}\noindent\textbf{US 5: Comparison with \textsc{neuron}~\cite{neuron}.} Lastly, we compare \textsc{lantern} with \textsc{neuron}~\cite{neuron}. In order to compare the full features of \textsc{lantern}, we \textit{integrate} \textsc{rule-lantern} and \textsc{neural-lantern} into a single framework for generating \textsc{nl} descriptions. Specifically, we track (\textsc{qep}, \textsc{nl} description) pairs viewed by each participants. By default, the \textsc{nl} description of each physical operator is generated using \textsc{rule-lantern}. Whenever an operator appeared more than a pre-defined \textit{frequency threshold} (\ie  5) in total in different \textsc{qep}s associated with a participant, \textsc{neural-lantern} is invoked to generate the description for the operator. 

Firstly, we ask the volunteers how well these two frameworks describe the query execution steps for \textsc{tpc-h} and \textsc{sdss} workloads. Figure~\ref{qsuppus} reports the results. \textsc{sdss} on SQLServer is not supported by \textsc{neuron} as it is tightly integrated with PostgreSQL and does not expose a declarative framework like \textsc{pool}. The translation rules for various operators of PostgreSQL are hardcoded in \textsc{neuron}. Consequently, even if we allow \textsc{neuron} to use \textsc{lantern}'s  parser for SQL Server to extract the operator tree from a given \textsc{qep}, none of the workloads of \textsc{sdss} is successfully translated as majority of operators of SQL Server have different names from those in PostgreSQL. Consequently, 41 out of 43 volunteers gave a score lower than 3 for \textsc{neuron}.  Secondly, we compare the boredom index of \textsc{neuron} and \textsc{lantern} for \textsc{tpc-h} on PostgreSQL. As shown in the last two rows of Table~\ref{tb:boredum}, the volunteers found the output of rule-based \textsc{neuron} more boring than \textsc{lantern}. Thirdly, \textsc{neuron} (resp. \textsc{lantern}) takes on avg.0.015 sec (resp. 0.172 sec) to generate a \textsc{nl} description. The avg. length of the descriptions is 188.136 (resp. 188.318) tokens.

\vspace{1ex}\noindent\textbf{US 6: Presentation models.} Recall that we use the presentation layer of \textsc{neuron}~\cite{neuron}  in \textsc{lantern}. Specifically, \textsc{nl} descriptions are presented in document-style text format. In this experiment, we  compare it with a \textit{visual tree-based \textsc{nl}} presentation format by \textit{integrating} the visual tree (Figure~\ref{fig:exp-qep}) with our \textsc{nl} description output. Specifically, the visual operator tree is shown by default and the \textsc{nl} description corresponding to each physical operator in the tree is added as an annotation to the corresponding node. A user can view the \textsc{nl} description of an operation by simply clicking on the corresponding node in the tree. An example is depicted in Figure~\ref{fig:vt-nl}. We ask the volunteers which of these two formats they prefer. Among  the 43 participants, 38 of them selected the document-style text format. Majority of learners are taking the database course for the first time. They mentioned that they chose the simple document-style presentation format as the visual tree-based \textsc{nl} format incurs a mental overhead  of integrating the \textsc{nl} descriptions associated with nodes and the sequence of steps depicted by the visual tree. In contrast, a text-based narration simply makes them read the text like a document (i.e., text book-style learning), which they are more familiar with.  

\vspace{0ex}
\section{Conclusions \& Future Work}\label{sec:conclusion}
The quest for high-quality techniques for natural language interaction with \textsc{rdbms} have witnessed a rejuvenation due to tremendous progress in deep learning and natural language processing. This paper takes a concrete step towards this grand vision by presenting a domain-oblivious framework called \textsc{lantern} that generates natural language descriptions of \textsc{qep}s to aide learners taking a database systems course. \textsc{lantern} provides a new paradigm of efficiently specifying \textsc{nl} descriptions of physical operators through a declarative interface, a rule-based technique that utilizes such specifications to translate a \textsc{qep} to its \textsc{nl} description, and a psychology-inspired deep learning-based framework  that adds diversity to the \textsc{nl} description in order to alleviate boredom among learners. We believe that \textsc{lantern}-generated descriptions of \textsc{qep}s complement its visual tree-based counterpart prevalent in commercial \textsc{rdbms}.  Our user study indeed demonstrates the effectiveness of \textsc{lantern} in facilitating comprehension of \textsc{qep}s among learners. As part of future work, we wish to explore techniques to facilitate \textsc{nl} interaction with a query optimizer to comprehend \textsc{qep} selection.

\vspace{0.5ex}\small{\textbf{Acknowledgments.} Sourav S Bhowmick and Shafiq Joty are supported by AcRF Tier-1 Grant 2018-T1-001-134. Hui Li is supported by National Natural Science Foundation of China (No. 61972309). We would like to thank Dr Patricia Chen from NUS (Dept of Psychology) for her advice on research related to habituation and boredom. We also thank Zheng Li from Xidian University for contributing to the implementation of \textsc{lantern} on SQL Server.

\bibliographystyle{abbrv}

\end{document}